\documentclass[%
twocolumn,
superscriptaddress,
 amsmath,
 amssymb,
 aps,
]{revtex4-1}
\usepackage{amsmath, latexsym, soul, color, type1cm,dsfont, float, fancyhdr}
\usepackage{graphicx}
\usepackage{ulem}
\usepackage{dcolumn}
\usepackage{bm}
\usepackage{color}
\usepackage{amsmath}
\usepackage{subfigure}
\usepackage[T1]{fontenc}
\usepackage{adjustbox}
\usepackage{graphicx}
\usepackage{dcolumn}
\usepackage{array}
\usepackage{bm}
\usepackage{amssymb}
\usepackage{amsmath}
\usepackage{subfigure}
\usepackage{physics}
\usepackage{sublabel}
\usepackage[utf8]{inputenc}

\usepackage{hyperref}
\usepackage{graphicx}
\usepackage{xcolor}
\hypersetup{colorlinks = true, linkcolor=blue, citecolor=blue, urlcolor=}
\begin{document}
\preprint{APS/123-QED}
\title{A simplified drift-diffusion model for pandemic propagation}
\author{Clara Bender}
\affiliation{Dept of Mechanical and Aerospace Engineering, University of Virginia, Charlottesville VA}
\author{Abhimanyu Ghosh}
\affiliation{Poolesville High School, Poolesville, MD}
\author{Hamed Vakili}
\affiliation{Dept of Physics, University of Virginia, Charlottesville VA}
\author{Preetam Ghosh}
\affiliation{Dept of Computer Science, Virginia Commonwealth University, Richmond VA}
\author{Avik W. Ghosh}
\affiliation{Department of Physics, University of Virginia, Charlottesville, Virginia 22903, USA}
\affiliation{School of Electrical and Computer Engineering, University of Virginia, Charlottesville, Virginia 22903, USA}
\date{\today}
\begin{abstract}
Predicting Pandemic evolution involves complex modeling challenges, often requiring detailed discrete mathematics executed on large volumes of epidemiological data. Differential equations have the advantage of offering smooth, well-behaved solutions that try to capture overall predictive trends and averages. We further simplify one of those equations, the SIR model, by offering quasi-analytical solutions and fitting functions that agree well with the numerics, as well as COVID-19 data across a few countries. The equations provide an elegant way to visualize the evolution, by mapping onto the dynamics of an overdamped classical particle moving in the SIR configuration space, drifting down gradient of a potential whose shape is set by the model and parameters in hand. We discuss potential sources of errors in our analysis and their growth over time, and map those uncertainties into a diffusive jitter that tends to push the particle away from its minimum. The combined physical understanding and analytical expressions offered by such an intuitive drift-diffusion model could be particularly useful in making policy decisions going forward.  
\end{abstract}
\maketitle
\section{Introduction}
Numerical modeling of pandemic propagation has a rich and varied history \cite{cov1,cov2,anil1,anil2,bertozzi,cleveland,cao,shakeel,gnanvi,lalmuanawmaa,roda,toda,tolles,vespignani}, ranging from Monte Carlo simulations that create histograms out of stochastic events, to Machine Learning/AI models trained on emerging data, curve fitting or structural models, graph theoretical approaches, to solving smooth differential equations such as  predator-prey (Lotka-Volterra) and Susceptible-Infected-Recovered (SIR) models  that simulate their average trends. Many of these models are highly detailed with several retro-fitted parameters. The ability of these models to predict accurate trends is often compromised by various sources of errors, as well as unpredictable uncertainties associated with geopolitics. What is critical to understand in this context are the simplified physical intuitions that may arise from these models, a minimal set of features and ways to capture them effectively, as well as the impact of various errors on their long term predictability.  

The purpose of this paper is multi-fold. 
\\\indent(a) We revisit the SIR model and relate its mathematical parameters with  key epidemiological constants, such as reproduction number $R_0$, herd immunity fraction $I_0$, incubation period $\tau$ and serial interval $S_I$ between events \cite{achaiah2}. The long-term behavior of its solutions can be expressed in terms of these parameters when they are time-independent constants. For instance, the long time single event fractional susceptibility $s^*$ can be expressed as a Lambert W function involving $R_0$, and further simplified to a power law over a range of $R_0$ values. 
\\\indent(b) We introduce an elegant physical picture underlying the dynamics, in particular, phase transitions associated with parameter tuning. The dynamics can be mapped onto the equation for an overdamped classical particle drifting down gradient of a potential profile $U(s)$, whose shape depends on the epidemiological constants as well as the specific model in hand. Damping makes the particle settle at the bottom instead of rolling through it, taking $s$ to its long-termed values $s^*$.\\
\indent(c) To make this solution quasi-analytical, especially for multi-events, we 
parametrically connect a simple model developed by Shur \cite{Shur2021} with the SIR model, relating causal inputs  (infection and recovery rates and their time-dependences) with observable consequences (stretch parameters, pandemic rise and fall times). \\
\indent(d) We apply this model across multiple countries over a much wider time period than originally explored in the Shur model.
In the process, we identify an inherent problem with a multiplicative model - namely, difficulty of fitting valleys without making the parameters unphysically large - an issue that argues towards an additive theory. A multiplicative model implicitly assumes independence of events, which makes it hard to pull out a new peak from a deep valley. We show that an analogous additive equation typically gives a larger $R^2$ fitting parameter.\\
\indent(e) Finally, we point out some of the sources of sensitivity in the model fitting parameters and their long-term impact on evolution. 
We introduce L\'angevin noise in the dynamics through both a Monte Carlo approach as well an equivalent drift-diffusion approach for the underlying, smooth probability distribution function (PDF), and interpret the evolution of the PDF by adding this diffusive jitter to the overdamped classical particle in an external potential. Just as a strong noise can kick a Brownian particle out of its global energy minimum to a shallower metastable state, strong uncertainty in parameters can limit the predictive time and lead to incorrect conclusions - not just evolutionary (continuously growing errors) but abrupt jumps (leaving one well and settling in another). We outline the impact of various uncertainties in the system that contribute to the diffusive spreading of the PDF, from initial reporting errors to parameter uncertainties to finite sampling size effects. 
\section{Simplifying the SIR Model}
Let us start by introducing the SIR model. 
The SIR model is a standard set of coupled differential equations used to study the evolution of an interacting population, such as one infected by a pandemic. The model is one of many approaches that have been invoked to study the spread of the COVID-19 pandemic. While elaborate Monte Carlo and AI models can delve into details, the model benefits from simplicity and the presence of a few, physically meaningful, lumped parameters. More importantly, the smoothness of the differential equations and their underlying solutions helps considerably in extracting quasi-analytical approximations (like we do here) and building physical intuition. The simplest version governing the time evolution of susceptible (S), infected (I) and recovered (R) population, where $S + I + R = N_p$ = constant (infected includes deceased population, in other words, the number of infections dead or alive), reads
\begin{figure}
\includegraphics[width=1.0\columnwidth]{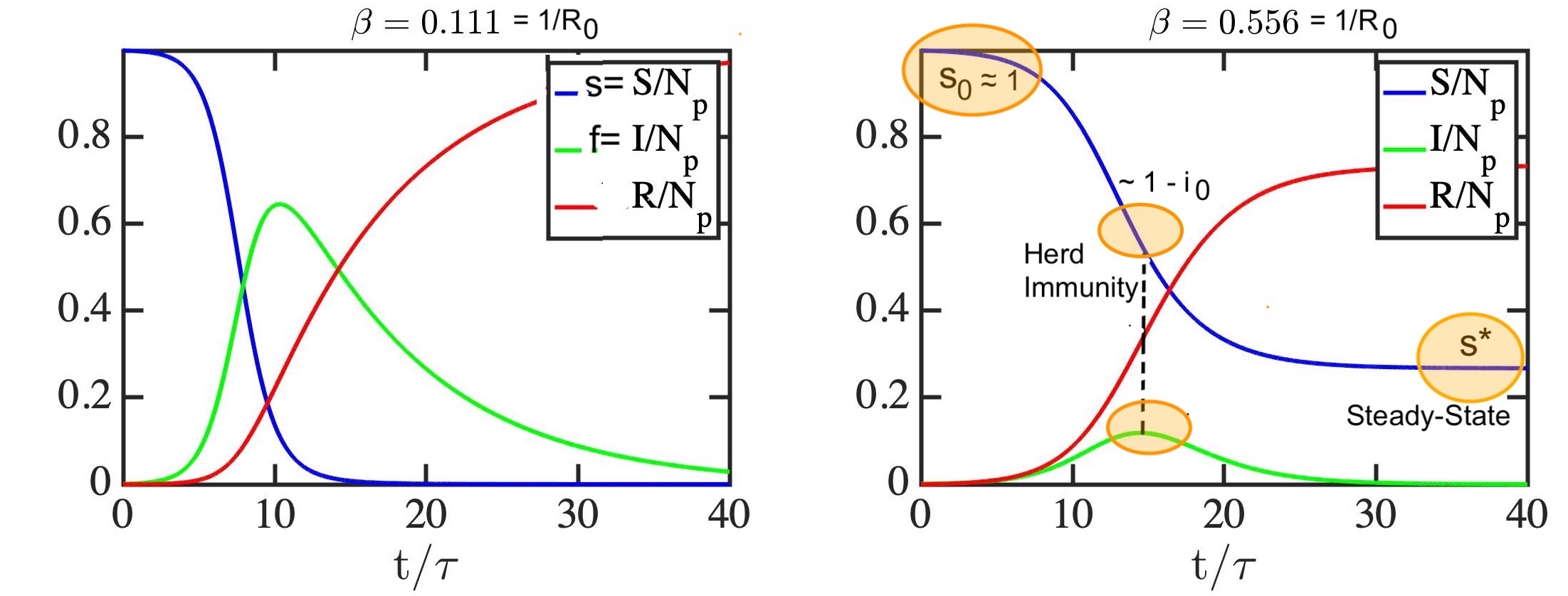}
\caption{\it{
The shape of SIR fractional populations plotted against a time axis scaled by $\tau_0 = 1/AN_p$  depends on a single parameter, $\beta = B/AN_p = 1/R_0$, $R_0$ being the reproduction number, and the initial condition $s_0 = S(0)/N_p \approx 1$. Here $N_p = 1000$ and $f_0 = 1/N_p$ starting with a single infected individual. The corresponding $R_0$ numbers for the two different $\beta$ value plots are 9 and 1.8. Key epidemiological metrics are outlined in orange circles. The herd immunity fraction $i_0$ is the fraction of infected plus recovered where the infected fraction peaks (meaning $s = 1-i_0$). $s^*$ is the steady-state value of $s$. In the simplest version of the SIR model, $s = \beta$  at herd immunity and $s^* \approx \beta^{2.5}$ at steady-state. This means herd immunity values are $1-\beta = 0.888$ (left) and $0.444$ (right), while $s^* \approx \beta^{2.5} = 0.004$ (left)  and $\sim 0.23$ (right).
}}
\label{f1}
\end{figure}
\begin{eqnarray}
\displaystyle \frac{dS}{dt} &=& -A SI \nonumber\\
\displaystyle \frac{dI}{dt} &=& (A S - B)I\nonumber\\
\displaystyle \frac{dR}{dt} &=& B I
\label{eqsir}
\end{eqnarray}
where $A$ is the average infection rate set by the strength of the interaction between infected and susceptible population, and $B$ is the average recovery rate. Here $N_p$ denotes the population of an infection cluster (one with spatially near-constant $A$, $B$ set across the population, set by a certain upper limit on the $|d(A,B)/(A,B)|$ fractional variation in parameter, with small interaction parameters $A$ between clusters in a multi-patch model). Both sets of parameters $A$ and $B$ are sensitively dependent on space (population density varies, governments act differently) and time (virus mutates, population grows herd immunity, medication improves, masking and quarantine policies evolve). By tweaking these parameters, we can build in effects such as different kinds of intervention \cite{chowdhury},  `stretching the curve', in other words, avoiding a breakdown of the healthcare system - in this case by imposing a simple threshold on $I$ that makes B plummet when I goes above a critical number $I_C$. The curves can then develop abrupt transitions or multiple modes. Later we show an example with a jump in $A$ leading to multiple peaks.

We can normalize the equation in terms of fractional populations,
$s = S/N_p$ and $f = I/N_p$ (the fractional recovered is unity minus the sum of $s$ and $f$). Dividing both sides by $N_p$, we get
\begin{eqnarray}
\dfrac{ds}{dt} &=& -\dfrac{sf}{\tau_0}\nonumber\\
\dfrac{df}{dt} &=& \dfrac{(s-\beta)f}{\tau_0}
\label{sirsimple}
\end{eqnarray}
where $\tau_0 = 1/AN_p$ and $\beta = B/AN_p$ are respectively, positive definite and positive semi-definite constants. 

For non-zero $\beta$, the set of equations has a fixed point at $f = 0$ where all the time-derivatives on the left vanish, giving a long-term asymptotic value of $f(t=\infty) = 0$ and $s(t=\infty) = s^*$. The intermediate values $s(t)$, $f(t)$ can be numerically obtained in Matlab using a straightforward ode23 solver, with an initial condition set by the initial fraction of infected population $f_0 \approx 1/N_p$, meaning we started the dynamics with a single infected population - (this in itself is an approximation, because a sizeable $(A,B)$ pair may only settle in after a few infected people are initiated). The corresponding initial condition on susceptible fraction is $s_0 = 1-f_0 = 1 - 1/N_p \approx 1$, since $N_p \gg 1$. 

Fig.~\ref{f1} shows the results for various $\beta = B/AN_p$ values. The quantity $AN_p = 1/\tau_0$ sets the overall incubation period (rise time), so that a plot of $f = I/N_p$ vs $t/\tau_0$ is controlled by the single parameter $\beta$ whose inverse $R_0 = 1/\beta$ is the reproduction number describing the average number of victims each infected person in turn infects.

From the second equation, we see that if at the outset $s_0 < \beta$, we have a negative slope in $f$ meaning the infection dwindles. In other words, for a recovery to infection ratio below a transition point $\beta^* = 1/R_0^* = s_0 \approx 1$, there is a non-zero steady state (i.e., $t \rightarrow \infty$) population $s^*$ of susceptible but uninfected population given by Eq.~\ref{eqL}, while the rest are all recovered. 

Finally, we have herd immunity $i_0$ that describes the fraction of the population that must be immunized
($1- s$, leaving a fraction $s$ susceptible) before the pandemic starts to dwindle. We see that this happens when $i_0 = 1 - \beta/s_0$. 
For the example in Fig.~\ref{f1}(b), we have $s_0 = 1-1/N_p = 0.999$ and $\beta = 0.556$ (contact number 
$R_0 = 1.79$), so that herd immunity sets in when the immunization rate is greater than $0.4434$, i.e., about $44 \%$ of the susceptible population is immunized, and the susceptible fraction has dropped to $0.556 = \beta/s_0$.


\section{Interpreting Pandemic Dynamics as an Overdamped Particle Drifting down a Potential}
We will now reinterpret the SIR equations to give them a physical picture.  If, for instance, we ignore recovery, then $s = 1 - f$ so that the equation for $f = I/N_p$ becomes a single variable one
\begin{equation}
    \displaystyle \frac{d{f}}{dt} = \frac{{f}(1-{f})}{\tau_0}, ~~~ \displaystyle \frac{d{s}}{dt} = -\frac{{s}(1-{s})}{\tau_0}, ~~~~ s + f = 1
\end{equation}
with ${f}_0 = 1/N_p \ll 1$ the initial fraction of infected  people. The solution is a modification of the well-known inverse Fermi-Dirac distribution 
$f = f_0/({f}_0+e^{-t/\tau_0})$ describing the equilibrium population of holes in a non-degenerate semiconductor (in neural net language it is the sigmoid/logistic function). It is notable that in the absence of any recovery and with an initial condition $f_0 > 0$, $f$ can only grow over time, meaning the long-term solution is $f = 1$, i.e., the entire population gets infected.  
We can provide an elegant physical interpretation of the evolutionary dynamics of the infected fraction $f$, if we interpret $f$ as a generalized configurational coordinate between 0 (uninfected) and 1 (fully infected). 
Since the total population does not change, the equation can be interpreted as a conservative picture of an overdamped classical particle whose distribution in time follows the down gradient of a potential 
\begin{eqnarray}
    \displaystyle \tau_0\frac{df}{dt} &=&  -\frac{\partial U(f)}{\partial f},  \nonumber\\
    U(f) &=&  -\dfrac{f^2(1-2f/3)}{2}
\end{eqnarray}
where $U(f)$ represents the potential, $0 < f < 1$. Here $df/dt$ is the speed of the particle, the inverse incubation time $\gamma = 1/\tau_0$ becomes its dynamic friction coefficient, and the acceleration term $md^2f/dt^2$ has dropped out as the particle has already reached terminal velocity where the frictional force $\tau_0 df/dt$ matches the driving force $F = -\partial U/\partial f$. 

\begin{figure}[ht!]
\includegraphics[width=0.848\columnwidth]{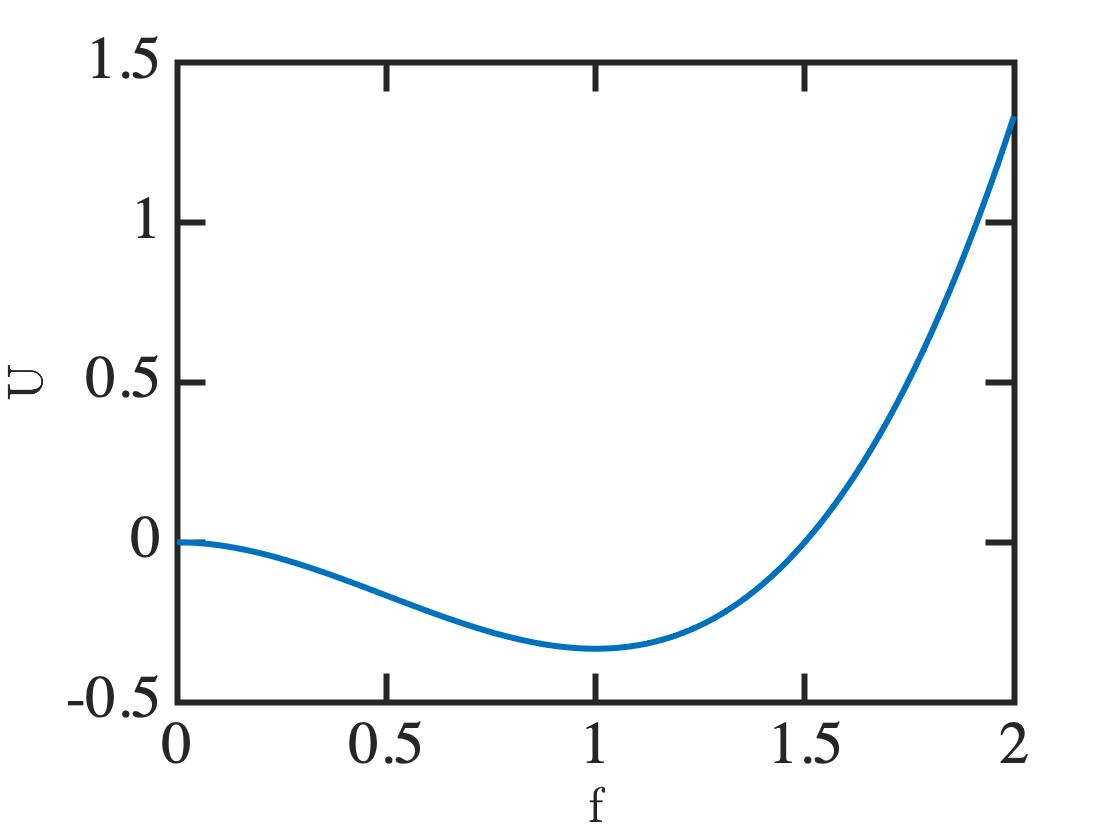}
\caption{\it{The evolution of the time-dependent probability distribution function (PDF) in the SIR model can be viewed as an overdamped particle moving in an equivalent potential plotted along the fractional infected space $f = I/N_p$ axis. This curve assumes no recovery ($B = \beta = 0$), which places the minimum energy and the steady-state at $f=1$ (all infected eventually).}}
\label{f2}
\end{figure}

Fig.~\ref{f2} shows the potential in question. It has a clear minimum at $f = 1$, meaning that in the absence of recovery the inevitable eventuality is for the entire population to get infected over time. 
\begin{figure}[ht!]
\includegraphics[width=0.48\columnwidth]{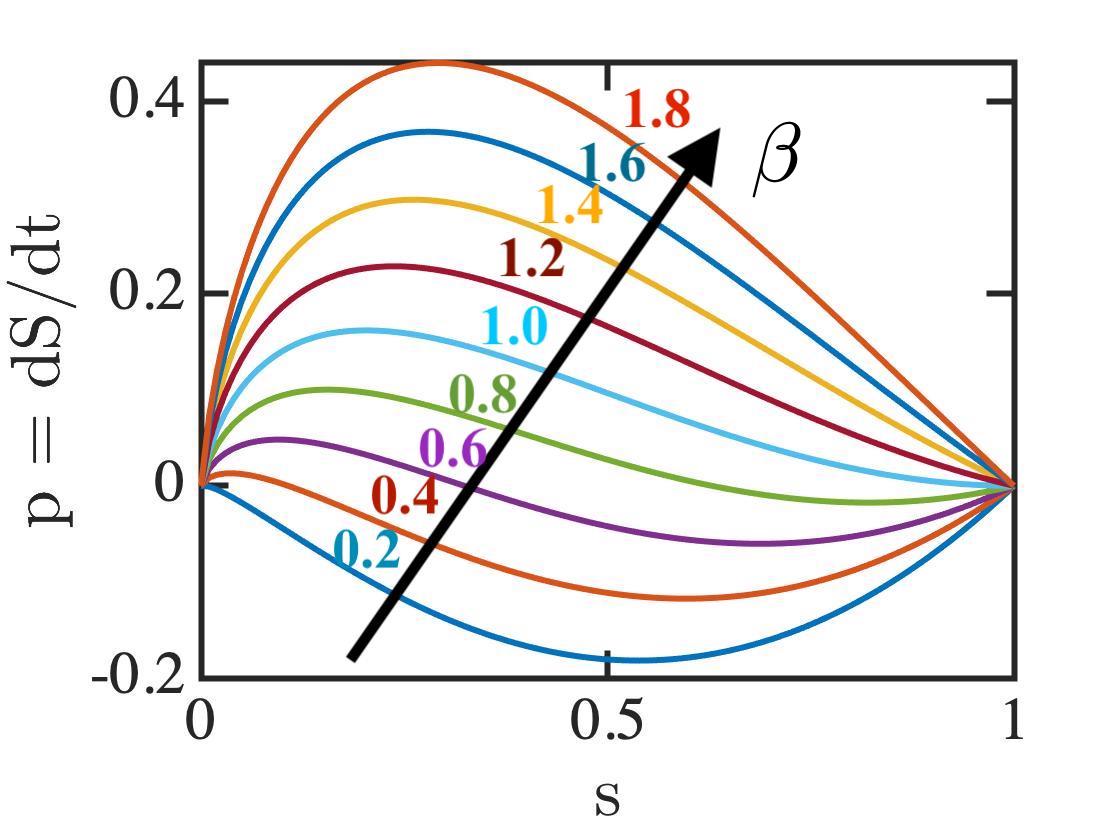}
\includegraphics[width=0.48\columnwidth]{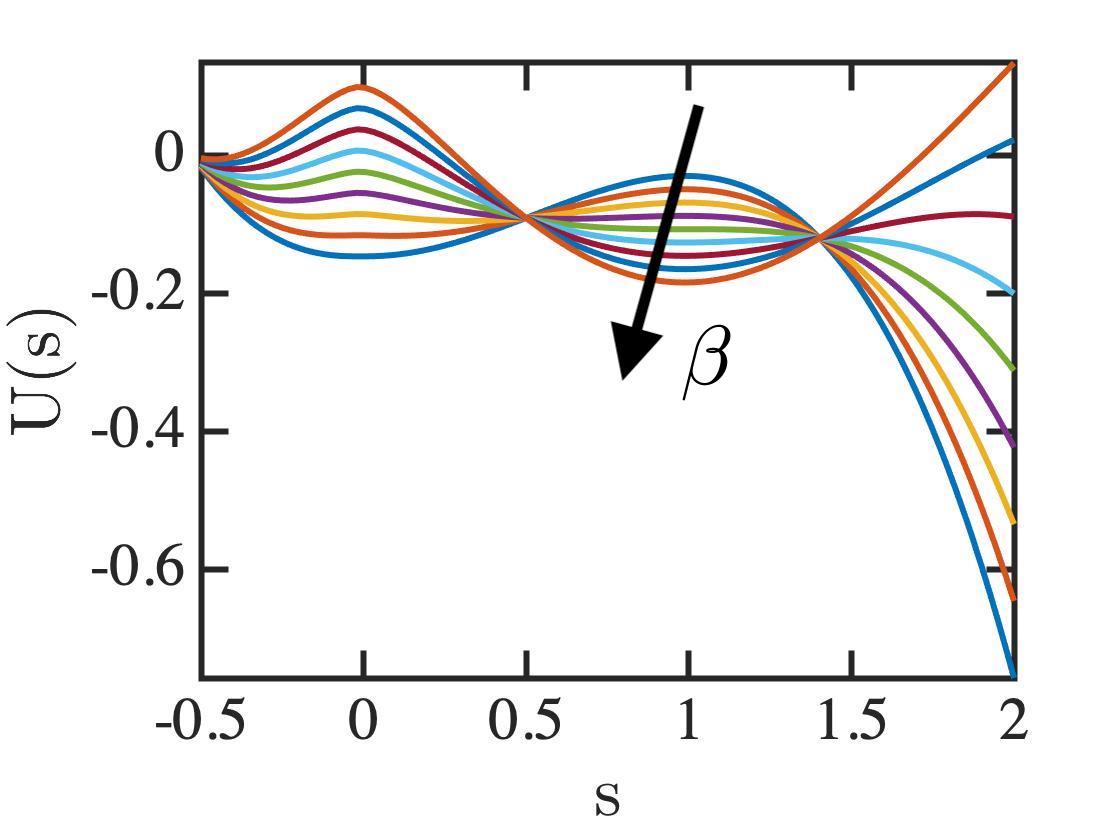}
\caption{\it{Varying the recovery rate $\beta$ (= inverse contact number) changes the speed of the $s$ evolution and the shape of the equivalent potential plotted vs $S$. With increasing $\beta$, the steady-state $s^*$ set by the potential minimum moves from 0 to 1. Note that the ground (zero) of the potential where all the curves meet is arbitrary. Here we set $U(s=-1/2) = 0$.}}
\label{f3}
\end{figure}
In presence of recovery $B$, the physics gets richer as we reach a steady state $s^*$.
Let us rewrite the differential equation for $s$ starting from Eq.~\ref{sirsimple}. 
We define $d^2s/dt^2 = pdp/ds$, where $p = ds/dt$, and use an integrating factor $1/s$ for $dp/dt$, in order to set up a differential equation in the ($s,p$) phase space. From there, we can get a differential equation involving the single variable $s$
\begin{eqnarray}
    \dfrac{d{s}}{dt} &=& -\dfrac{{s}(1-{s})}{\tau_0} - \beta \left(\dfrac{s\ln{s}}{\tau_0}\right) = -\dfrac{1}{\tau_0}\dfrac{\partial U(s)}{\partial s}\nonumber\\
    U_{SIR}(s) &=& \dfrac{s^2(1-2s/3)}{2} + \beta\left[\dfrac{s^2(2\ln{s}-1)}{4}\right]
    \label{eqmotion}
\end{eqnarray}
upto an additive constant (similar to the choice of a ground in an electrical potential), where once again $\beta = B/AN_p = 1/R_0$, $R_0$ being the reproduction number. 
The evolution of the potential $U_{SIR}(s)$ for various recovery to infection rate ratios $B/A$, and the corresponding force $F_s = -\partial U/\partial s = \tau_0 ds/dt$ are shown in Fig. ~\ref{f3}.

\begin{figure}[!ht]
\includegraphics[width=.47\textwidth]{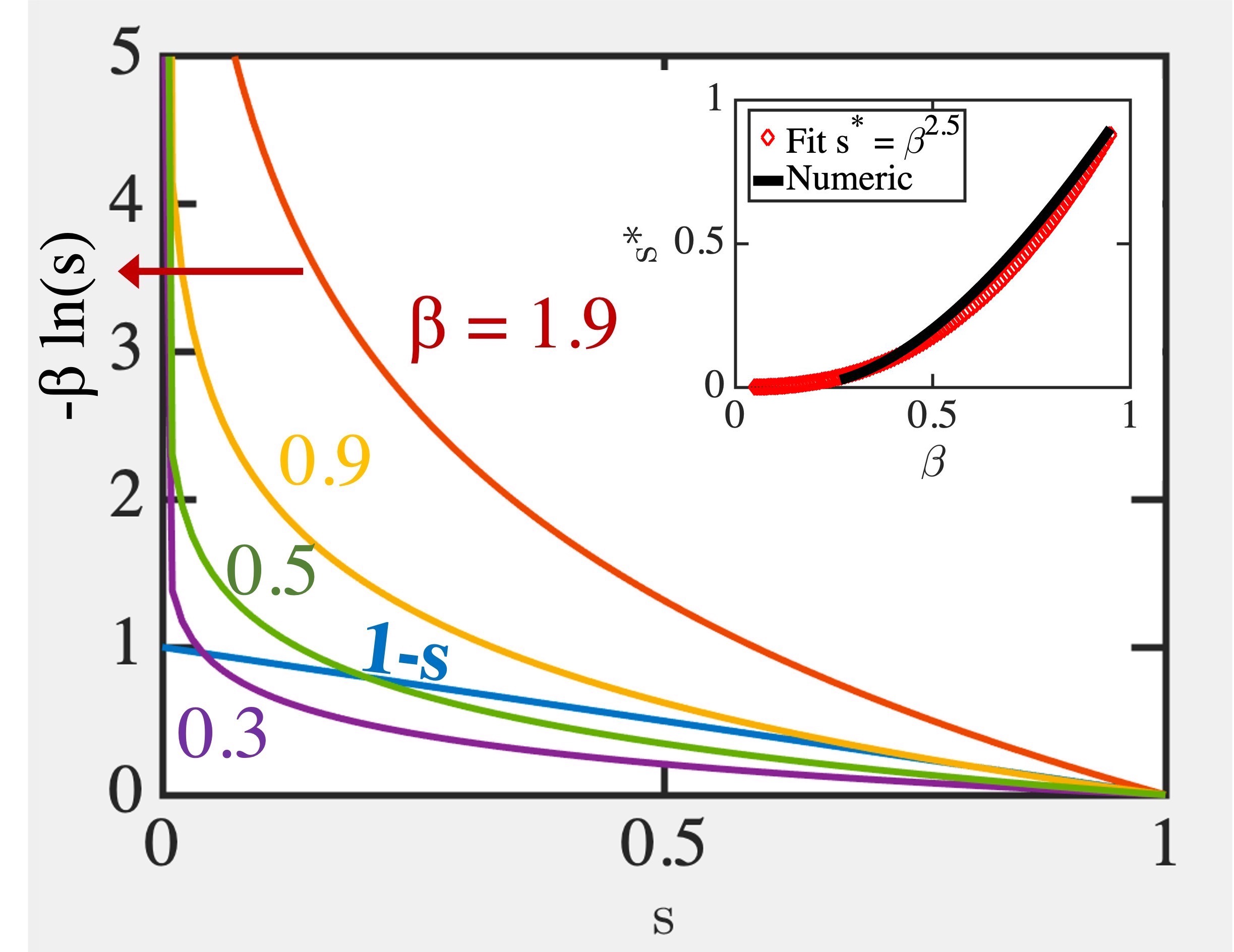}
\caption{\it{Plots of the transcendental equation (Eq.~\ref{eqt}) for $s^*$, the steady-state susceptible population, given by the intersection of a straight line and a $\beta$-dependent family of curves. While the solution involves Lambert functions (Eq.~\ref{eqL}), an approximate solution can be postulated as a power law (inset). } }
\label{fna}
\end{figure}

It is prudent at this stage to find the fixed points of the equation where $ds/dt = 0$. We get two fixed points, $s=0$ and $s=1$ for $\beta > 1$, and more interestingly for $\beta < 1$, an intermediate number $s^*$ satisfying 
\begin{equation}
    1-s^* = -\beta\ln{s^*}
    \label{eqt}
\end{equation}
The right side of this transcendental equation starts from 0 at $s = 1$ with initial slope $-\beta$ and rises monotonically with decreasing $s$ (Fig.~\ref{fna}), so it can only intersect the left part of the equation if its slope is smaller in absolute magnitude than the fixed slope $-1$ of the left side, i.e., $\beta < 1$. This means there is a steady-state fraction of recovered population as well, set at $1-s^*$. The exact solution is given by
\begin{equation}
    {s^*} = -\beta
    W\Bigl(\dfrac{-e^{-1/\beta}}{\beta}\Bigr)
    \label{eqL}
\end{equation}
where $W$ is the Lambert W function (the inverse of $xe^x$), in this case extended along the negative branch \cite{WL,WL2}.
Over a reasonable range of $\beta$ values shown above, the solution roughly follows a power law $s^* \approx \beta^{5/2}$, for $\beta < 1$ (inset in Fig.~\ref{fna}). Indeed, using the parameters from Fig.~\ref{f1}, we see that for $\beta = 0.556$, $s^*N_0 = S_0 = 268$, while the approximation $\beta^{2.5} = 230$, in decent agreement with the plot. For $\beta = 0.111$, $S_0 = 0.1$ ie it reaches zero. The overall message is simple - that reducing $\beta$ (increasing the $R_0$ number) reduces the fraction of the population that stays uninfected over time. \\
\indent The equation above accords no analytical solution for $s(t)$ and thus $f(t)$ beyond the fixed point. However, it has the general shape of a Fermi-Dirac distribution and can be solved to arbitrary accuracy numerically, using the evolution of a classical overdamped particle in an evolving potential. 
The situation can be complicated by the emergence of multiple infection and recovery events, a complex geopolitical situation with evolving knowledge, healthcare, government decisions, test taking etc, which can in their totality make parameter estimation necessary for prediction near impossible beyond a certain time frame (not to mention that severe nonlinearity can make prediction very short ranged even if the parameters were somehow known to reasonable accuracy). The focus therefore is to look at the generic structure of the solutions, and qualitative wisdom arising from them. \\
\indent One problem with this equation is the fact that at large $t$, $I = 0$ which means everyone is either uninfected or fully recovered. This is not consistent with multiple events, and is in fact a consequence of the exponential drop in $I$, especially around the fixed point of $s$ i.e., $df/dt \approx -(s^*-\beta)f/\tau_0$. 
To counter this drop, we can make $\tau$ a stretchable time $\tau = \tau_0 + \alpha t$, which gives a slower decay $f \sim 1/t$ for long times at which point a change in $A$ can allow a re-emergence to occur. 
Indeed, we can modify our SIR equations to accomodate the stretch parameter $\alpha$, using a revision of the time axis that stretches from linear to logarithmic
\begin{eqnarray}
    t^\prime &=& \dfrac{\tau_0}{\alpha}\ln{\left(1 + \dfrac{\alpha t}{\tau_0}\right)}\nonumber\\
   &=& \begin{cases}t, ~~ \alpha t\ll \tau_0 \\
    \\
    \dfrac{\tau_0}{\alpha}\ln{\dfrac{\alpha t}{\tau_0}}, ~~ \alpha t \gg \tau_0\end{cases}
\end{eqnarray}
whereupon we get 
\begin{eqnarray}
    \displaystyle \frac{ds}{dt^\prime} &=&   -\dfrac{1}{\tau_0}\frac{\partial U(s)}{\partial s} \nonumber\\
    U_{SIR}(s) &=& \dfrac{s^2(1-2s/3)}{2} + \beta\left[\dfrac{s^2(2\ln{s}-1)}{4}\right]~~
\end{eqnarray}
In the analysis of a simple phenomenological model that we will borrow from \cite{Shur2021}, the infected and recovered are both captured within one count $f + r = 1-s$. As can be seen in the adjoining plots, this fraction amounts to a stretch out of the susceptible population from $\sim 0$ to a finite value $1-s^*$, created by the recovery rate that endows the Shur analysis with a phenomenological stretch parameter $\alpha$:
\begin{equation}
    \dfrac{df}{dt} = \dfrac{f_0e^{t/\tau}}{(f_0 e^{t/\tau}+1)^2\tau}, ~~ \tau = \tau_0 + \alpha t
    \label{shuralpha}
\end{equation}
where $f = 1-s$, and $\alpha$ is adjusted to the peak position $t_m$ as, 
\begin{equation}
    \alpha = -\left[\dfrac{\tau_0}{t_m} + \dfrac{1}{\ln{\dfrac{1}{f_0}}}\right]
\end{equation}

\begin{figure}[!ht]
\includegraphics[width=0.5\textwidth]{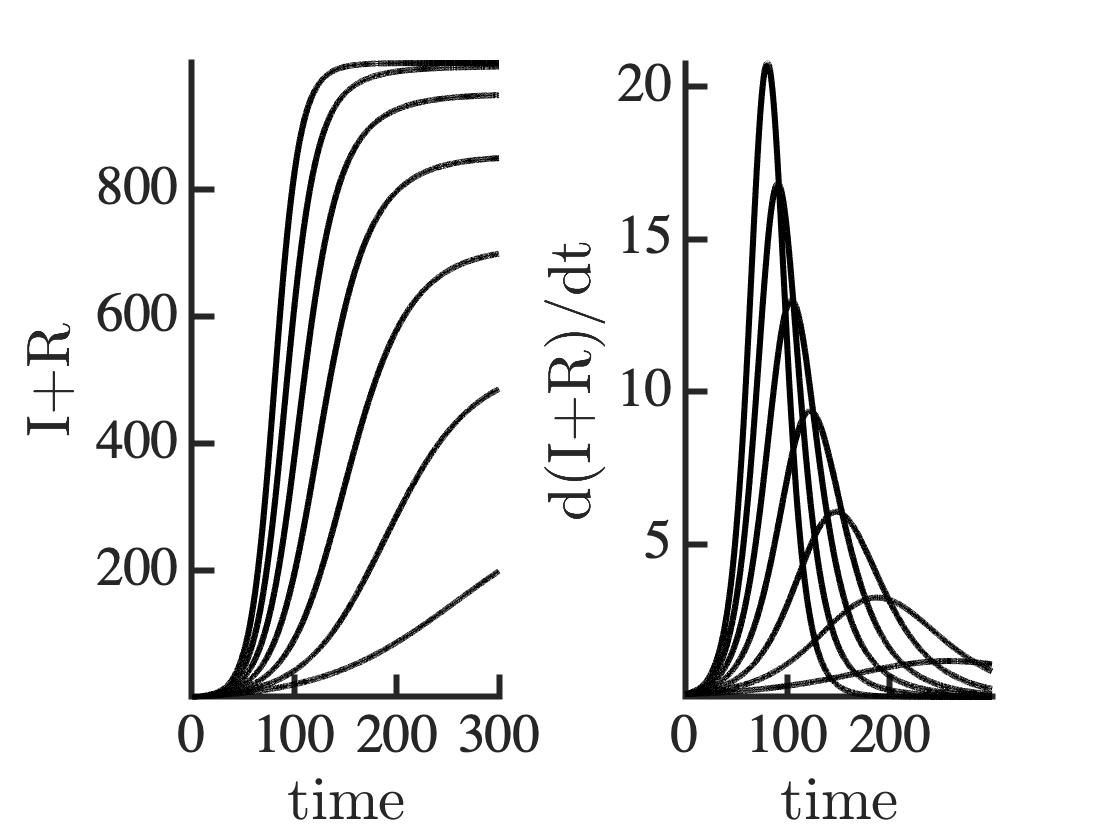}
    \caption{\it{An effective stretch parameter ($\alpha$ in the treatment by Shur) arises for the non-susceptible population if we vary the recovery rate $\beta \propto B$ in the SIR model. Time is measured in days.}}
\label{stretch}
\end{figure}

\begin{figure}[!ht]
\includegraphics[width=0.5\textwidth]{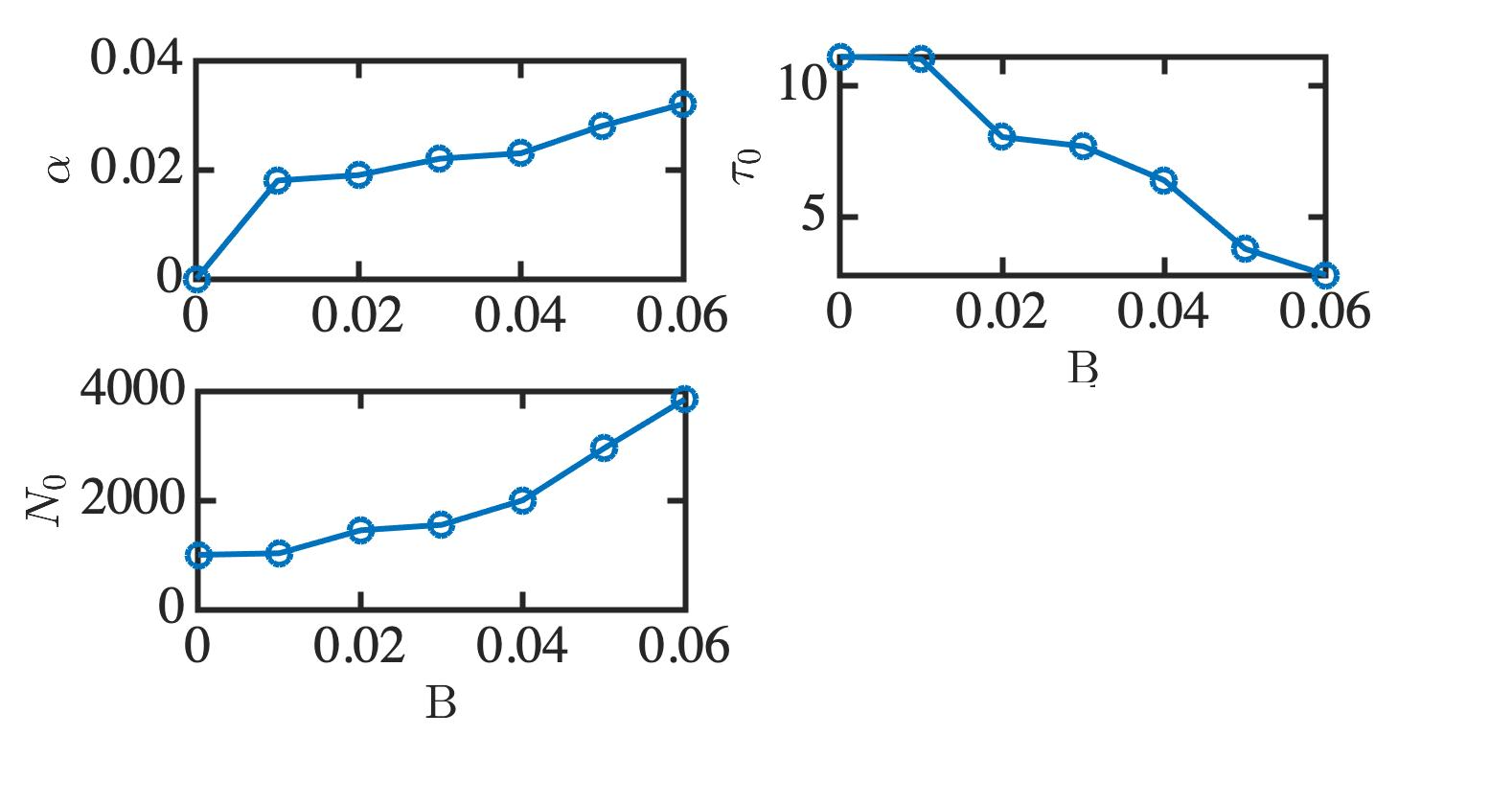}
    \caption{\it{Dependence of the Shur parameters $\alpha$ (stretch), $\tau_0$ (initial rise time) and $N_p$ (effective susceptible population) on the SIR recovery parameter $B$, extracted by fitting Fig.~\ref{stretch} with Eq.~\ref{shuralpha}.}}
\label{stretch2}
\end{figure}

Fig.~\ref{stretch} plots the SIR evaluated infected + recovered (i.e., once infected) population for various relative recovery rates $\beta$,
and shows that the effect of increasing recovery is to endow the bell curve with a larger stretch out character. We fitted the numerical SIR results with the Shur single peak equation (Eq.~\ref{shuralpha}), and show how the Shur parameters relate to the SIR parameters (Fig.~\ref{stretch2}). 
We see that the 
the extracted stretch parameter $\alpha$ which increases roughly logarithmically with $B$ , the corresponding reduction in initial rise time $\tau_0$ (to keep the area under the curve meaningful), and the effective susceptible population at $t=0$, $N_0 \approx N_p = 1/f_0$, a fitting parameter setting the maximum height of the $d(I+R)/dt$ curve, with $N_0$ also varying weakly with $B$.  
\section{Multiple Events}

\begin{figure}[!ht]
\includegraphics[width=.5\textwidth]{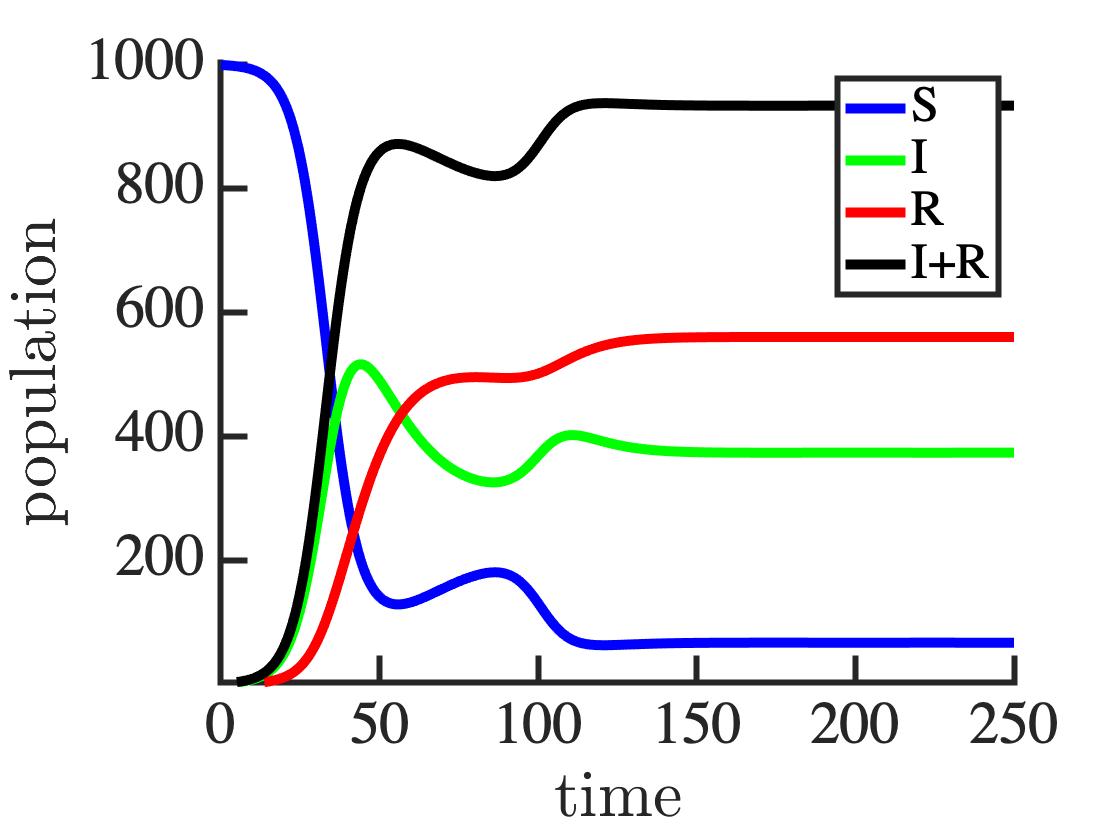}
\includegraphics[width=.5\textwidth]{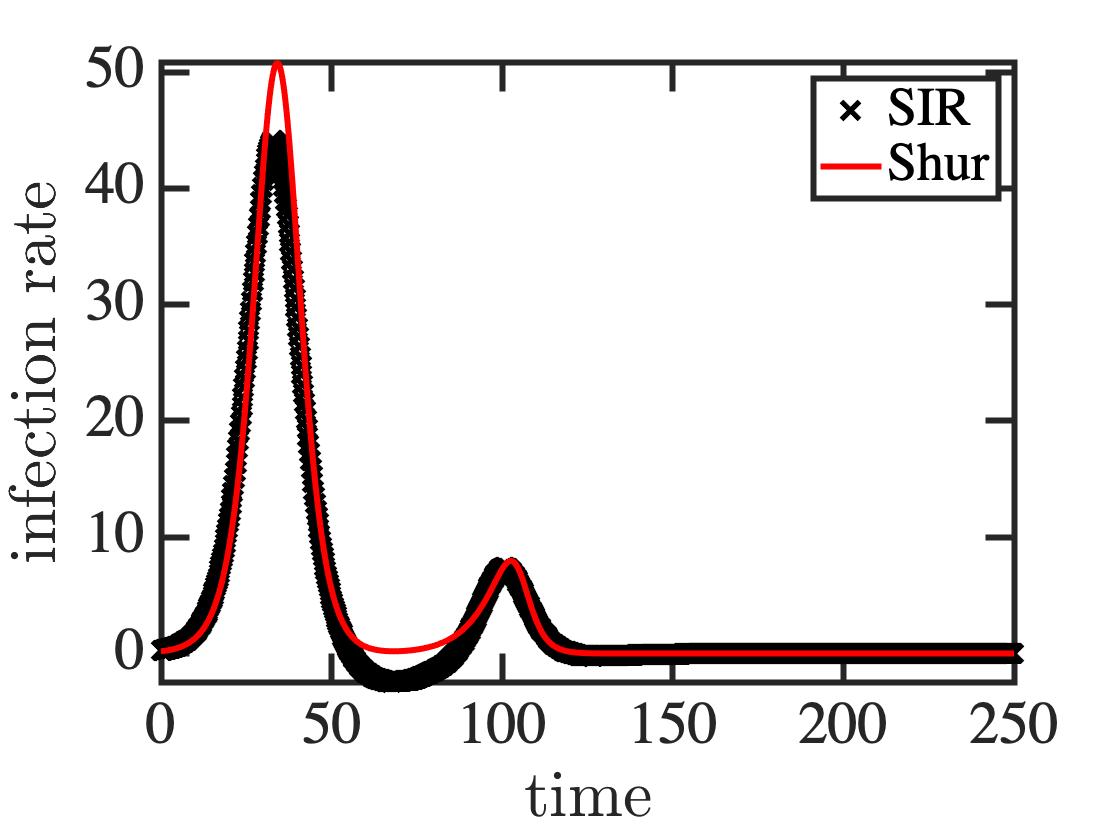}
    \caption{\it{The emergence of multi-peaks in the Shur model from a thresholded jump in infection rate (Eq.~\ref{eqthr}) in the SIR model. Parameters are $A_0=0.25$, $\beta_1 = 0.5$, $\beta_2 = 0$,  $N_0 = 1000$, $\tau_b = 5$, $t_{b1} = t_{b2} = 100$, $\alpha = -0.01$, $N_0 = 1000$, and $B = 0.05$. We also introduced a parameter $C = 1/30$ corresponding to an SIRS treatment where part of the recovered population is reinserted into the susceptible population.}}
\label{2peak}
\end{figure}

\begin{figure*}[!htbp]
\includegraphics[width=.47\textwidth]{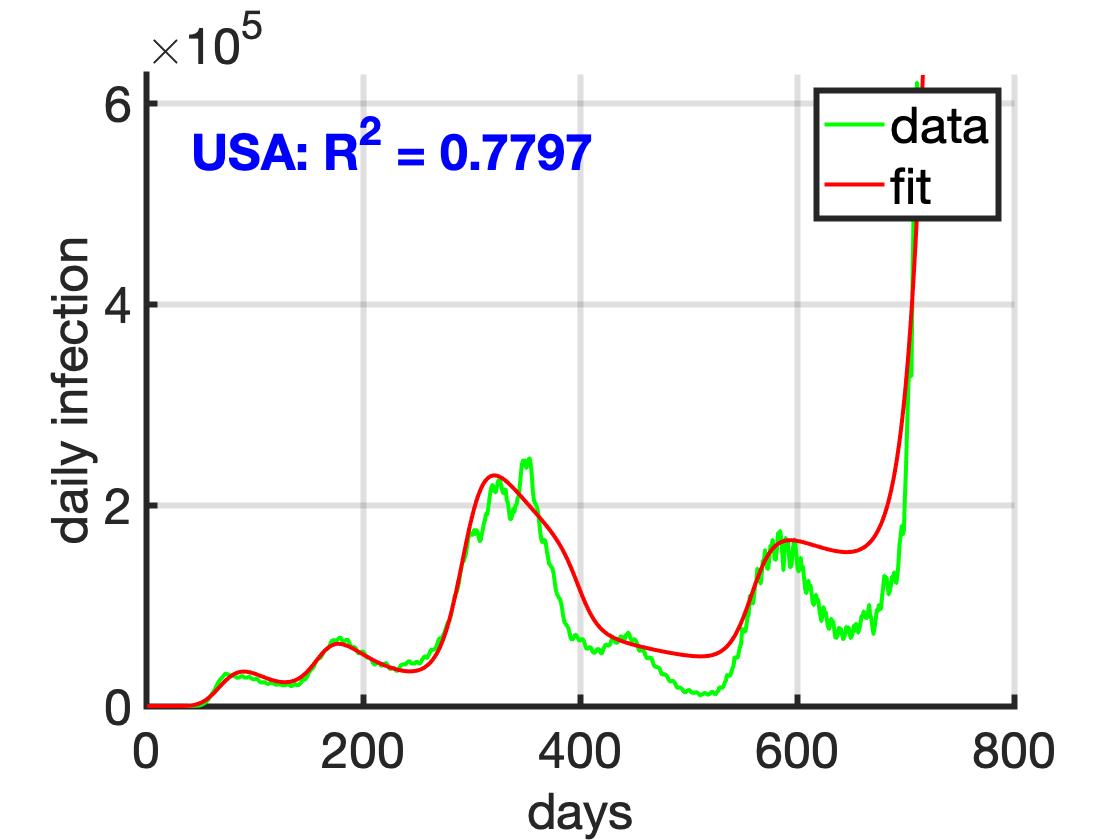}
\includegraphics[width=.47\textwidth]{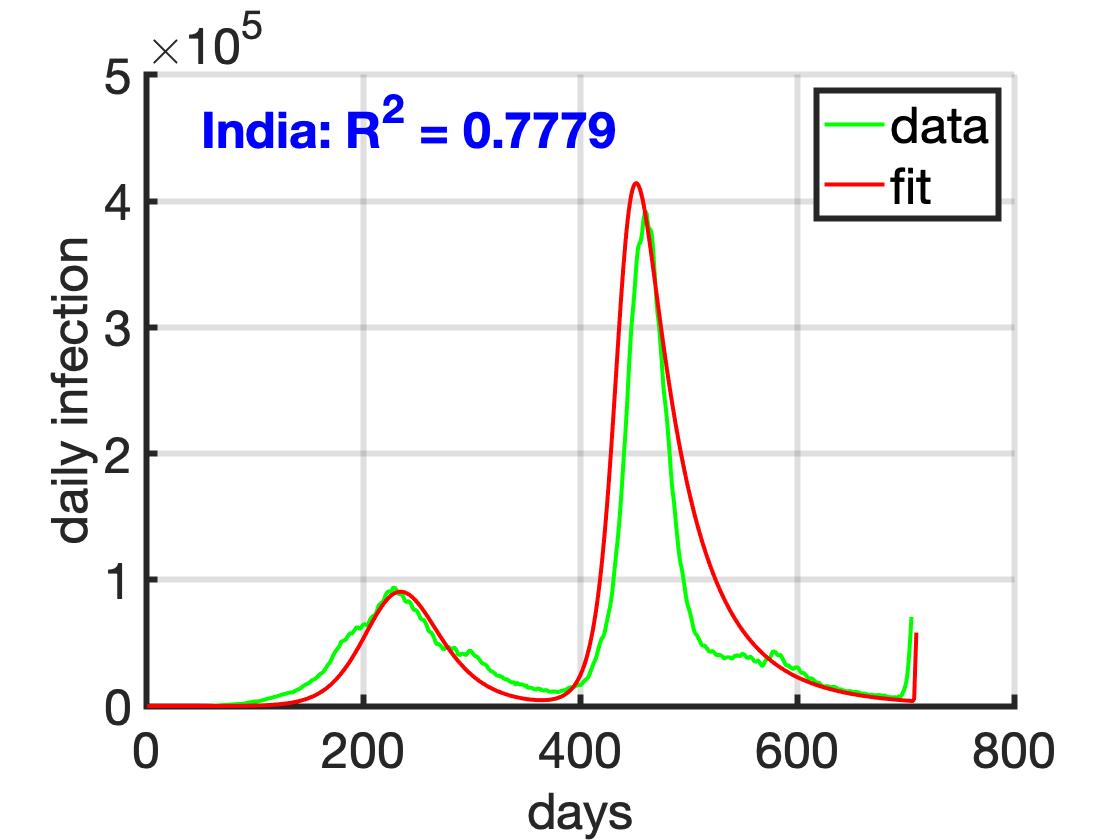}
\includegraphics[width=.47\textwidth]{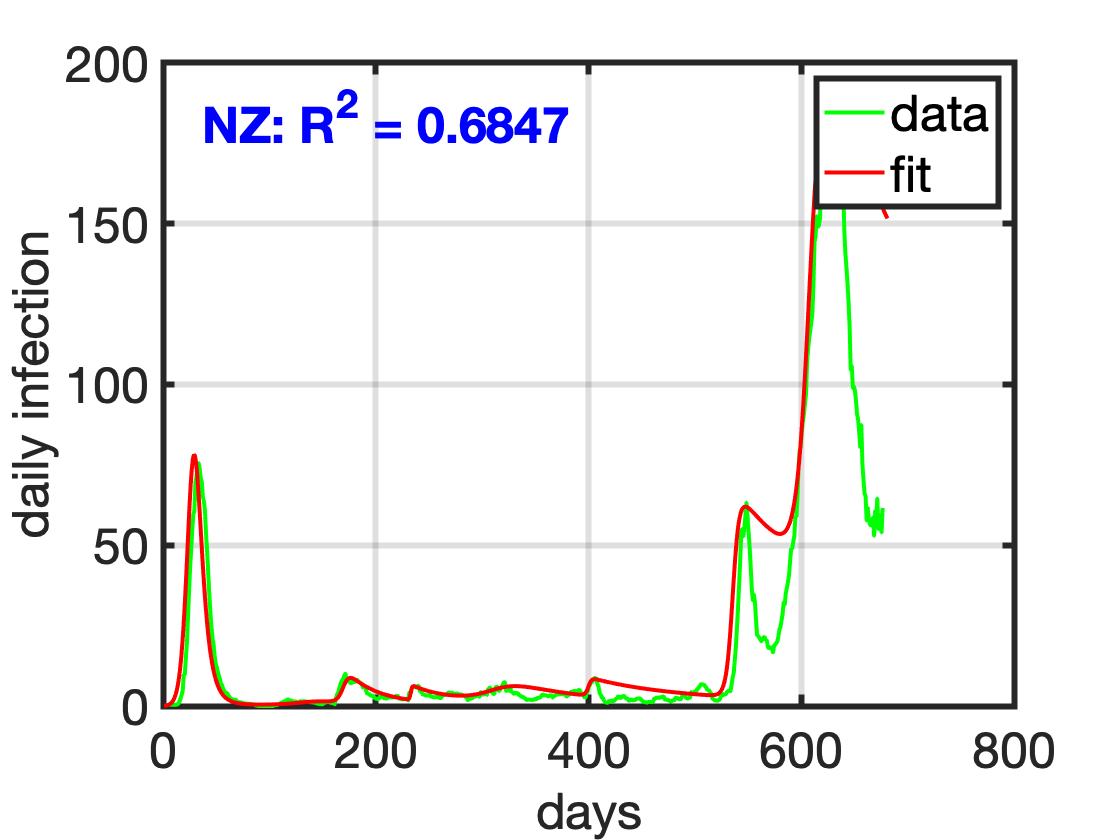}
\includegraphics[width=.47\textwidth]{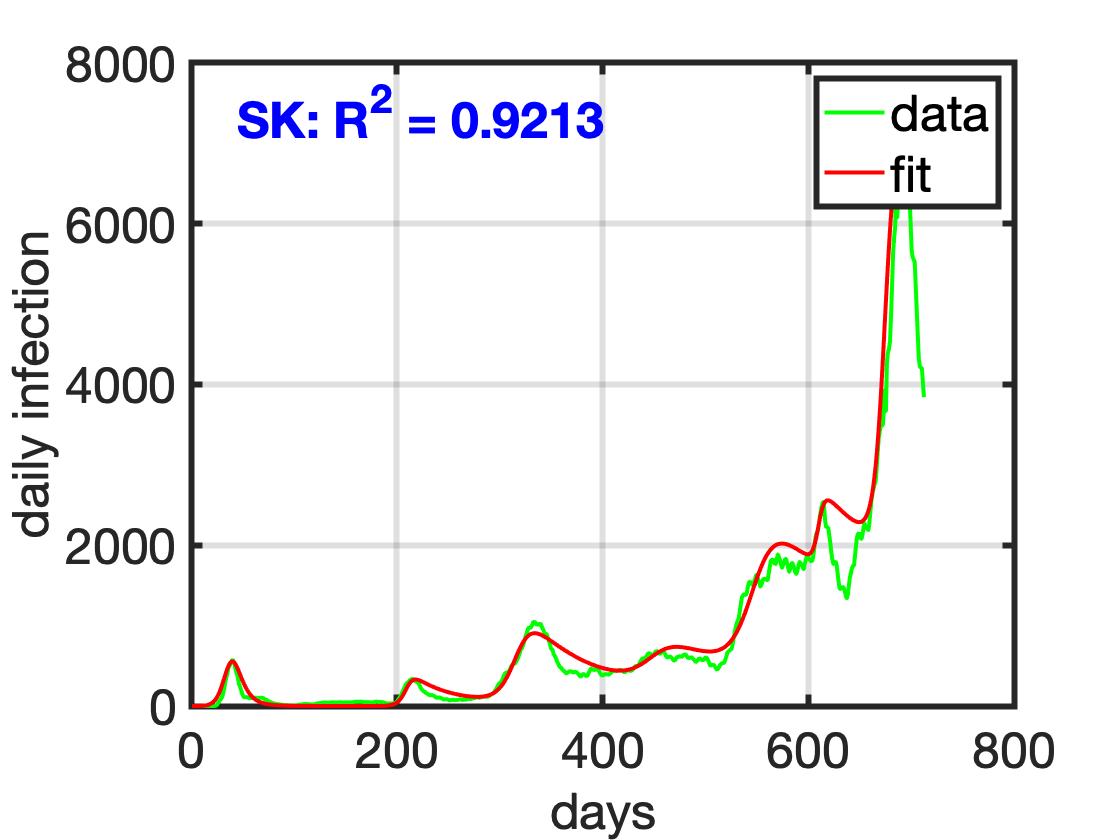}
    \caption{\it{Fits of infection data (source \cite{owid}), smoothened over a sliding interval of 10 days (green), with a fit based on the simplified multiplication Eq.~\ref{shurM} (red), with parameters listed in Table \ref{tablet}. The $\beta_i$ values corresponding to mitigation or demitigation are sensitive to the fits of the intervening valleys, becoming unphysically large due to the employment of a multiplicative (i.e., independent events) model (Eq.~\ref{shurM}) rather than an additive one (Eq.~\ref{eqadd2}). The fitted $R^2$ values are shown on the figures.}}
\label{fig2}
\end{figure*}

To accommodate multiple resurgence events, Shur's paper multiplies the analytical first peak result with other fermi/antifermi distributions piece-meal, keeping in mind that the final answer is not the sum of separate peak distributions, but independent products (meaning that separate fits of each peak, as conventional in Lorentzian fits to peaked data, will not work). We start with the following expression from Shur with phenomenological mitigation parameters $\beta_i$, peak times $t_{\beta i}$ and peak widths $\tau_{\beta i}$. 
\begin{equation}
    \dfrac{df}{dt} = \dfrac{f_0e^{\displaystyle t/\tau}}{\tau\left(f_0e^{\displaystyle t/\tau} + 1\right)^2} \times \prod_{n=1}^N \left(1 - \dfrac{\beta_n}{e^{\displaystyle (t-t_{\beta_n})/\tau_{\beta_n}}+1}\right)
    \label{shurM}
\end{equation}
where $\beta_n$ describes mitigation when positive (demitigation when negative) for the $n$th event, $t_{\beta_n}$ describes the onset of the event (roughly turning on at $t_{\beta_n} - 3\tau_{\beta_n}$) and $\tau_{\beta_n}$ is the recovery time over which the event persists. 

In keeping with the tone of this paper, let us try to justify this fitting form from the SIR equation. We can capture the same physics numerically with the SIR model brute force, by assuming in Eq.~\ref{eqsir}
a time-dependent infection rate 
\begin{equation}
    A(t) = A_0\left[1 + \dfrac{\beta_0}{1 + e^{\displaystyle -(t-t_\beta)/\tau_\beta}} \right]
= \begin{cases}
A_0, ~~t \ll t_\beta \\A_0(1+\beta_0), ~~t \gg t_\beta
\end{cases}
\label{eqthr}
\end{equation}
A good agreement between the SIR result with this varying $A$ and the Shur multi-peak equation above is showcased in Fig.~\ref{2peak}, correlating the parameters $\beta_0$, $\tau_\beta$ and $t_\beta$ in Eq.~\ref{eqthr} with the first mitigation peak parameters $\beta_1$, $t_{\beta_1}$, $\tau_{\beta_1}$ in Eq.~\ref{shurM}.   The interpeak separation $t_{\beta_{i+1}}-t_{\beta_i}$ is related to the serial interval $S_I$ between events. 

The original work introducing these equations worked with data fits over only one or two peaks across a short time period. We show now that this can be extended across much larger time scales in multiple populations. However, there are some prices to pay. As peak heights vary substantially, the $\beta_n$ values are sometimes much bigger (Table ~\ref{tablet}) than originally proposed (10s of thousands instead of between -1 to +3). They are on the one hand restrictive (small adjustments to even such large $\beta$ values change the later peaks substantially) and on the other hand sensitive to other parameters, primarily the stretch function $\alpha$. This is not altogether unexpected. The job of $\alpha$ is to sustain a background infected population that allows resurgence down the road (i.e., it creates a floor that the later peaks ride on). This makes it hard to remove any stretch features out of later peaks which necessarily become asymmetric
and makes it hard to capture deep valleys in the data. It also depends sensitively on the floor value that the first peak subsides to - while data on the floor is harder to gather, its magnitude can affect subsequent parameter values sensitively. Simply put, we need a large negative $\beta$ to pull a peak out of a very low valley, so errors in estimating the valley floor affect $\beta$ values quite sensitively. 

Fig.~\ref{fig2} and the accompanying table show an attempted fit for data \cite{owid} across multiple countries over several months. 
Let us briefly discuss the fitting protocol, as suggested by Shur \cite{Shur2021}. Alternate methods for SIR fits exist \cite{fit}, \cite{fit2}. We fit the rise time of the first peak with $\tau_0$ and its height with $f_0$. The peak position $t_m \approx \tau_0\ln{1/f_0}/(1-\alpha \ln{1/f_0})$ then gives us the stretch parameter $\alpha$. For subsequent peaks, the onset of a rise is roughly $t_\beta - 3\tau_\beta$, the peak width $\sim \tau_\beta$ and the height controlled by $\beta$ itself (a positive $\beta$ gives a drop while a negative $\beta$, seen commonly here, gives a rise). 
\\\\
\begin{table}
\begin{math}
\begin{array}{ccccc}
\hline
& USA & India & New Zealand & S Korea\\
\hline
\tau_0 & 0.02 & 9.7 & 2.2 & 2.2\\
f_0^{-1} & 700440 & 5510500 & 1000 & 8000\\
\alpha & 0.055 & 0.0234 & 0.0347 & 0.0347\\
\beta_1 & -10 & -1700 & -1116 & -28000\\
\tau_{\beta_1} & 10 & 10 & 3 & 3.8\\
t_{\beta_1} & 163 & 146 & 535.8 & 210\\
\beta_2 & -15.5 & -15 & -11 & -25\\
\tau_{\beta_2} & 11 & 2.8 & 6 & 8\\
t_{\beta_2} & 295 & 709 & 170 & 318\\
\beta_3 & 0.5 & N/A & -2.5 & -2.5\\
\tau_{\beta_3} & 9 & N/A & 0.6 & 12\\
t_{\beta_3} & 400 & N/A & 233 & 450\\
\beta_4 & -3.7 & N/A & -6 & -4.5\\
\tau_{\beta_4} & 10.8 & N/A & 12 & 11\\
t_{\beta_4} & 599.4 & N/A & 315 & 548\\
\beta_5 & -58.7 & N/A & -1.8 & -0.55\\
\tau_{\beta_5} & 15 & N/A & 1.5 & 3\\
t_{\beta_5} & 755.5 & N/A & 400 & 610\\
\beta_6 & N/A & N/A & -23 & -3\\
\tau_{\beta_6} & N/A & N/A & 3 & 5.5\\
t_{\beta_6} & N/A & N/A & 535.8 & 675\\
\beta_7 & N/A & N/A & -4 & N/A\\
\tau_{\beta_7} & N/A & N/A & 6 & N/A\\
t_{\beta_7} & N/A & N/A & 608 & N/A\\
\hline
\end{array}
\end{math}
\caption{Fitted Shur parameters (Eq.~\ref{shurM}) with OWID COVID-19 data \cite{owid}, qualitatively consistent with SIR numerical results with Eq.~\ref{eqthr}. Some of the mitigation parameters are exceptionally high in order to pull a following peak out of a deep valley.}
\label{tablet}
\end{table}
The goodness of fit can be quantified by the $R^2$ number listed on the figures. For a fitting function $z(t)$ (the phenomenological equation) compared to a target function $y(t)$ (the smoothened data), the fitting equation is given by
\begin{equation}
R^2 = 1 - \dfrac{\sum_i(z_i-y_i)^2}{\sum_i(y_i-\langle y\rangle)^2}
\label{eqadd2}
\end{equation}
where $\langle y\rangle$ is the time-averaged value of $y(t)$. Note that for truly bad fits where the predicted regression curve $z$ departs further from $y$ than does the mean, $R^2$ can in fact be negative; however, for a reasonable fit we expect it to lie between 0 and 1, and venture closer to zero as the fit gets better and better. Also close to near zero values, the denominator could numerically vanish faster than the numerator, so we will need to manually prune any Matlab outputs with NaN near zero. 

It is worth emphasizing that in spectroscopic analyses, fitting functions for multipeaked experimental data often decomposes naturally into {\it{sums}} of Lorentzians. Such a sum in effect allows us to fit the peaks easily, including the intervening valleys. A {\it{multiplicative}} model, in effect, treats the probabilities independently, which becomes a problem because the initial value for each peak is set by the valley floor (and thus the stretch parameter) of the first peak. An example of such fitting anomaly is seen from the data table. The fitted populations $N_p = f_0^{-1}$ follow the expected sequence across the countries. The $\beta_1$ values for India, South Korea and New Zealand are very high compared to the US, suggesting an aggressive initial mitigation strategy (quarantine, masking). However, the exact number is probably unphysical, as small variations in the fitted valley can alter $\beta$ in a hyper-sensitive fashion. 

\begin{figure*}[t!]
\includegraphics[width=.47\textwidth]{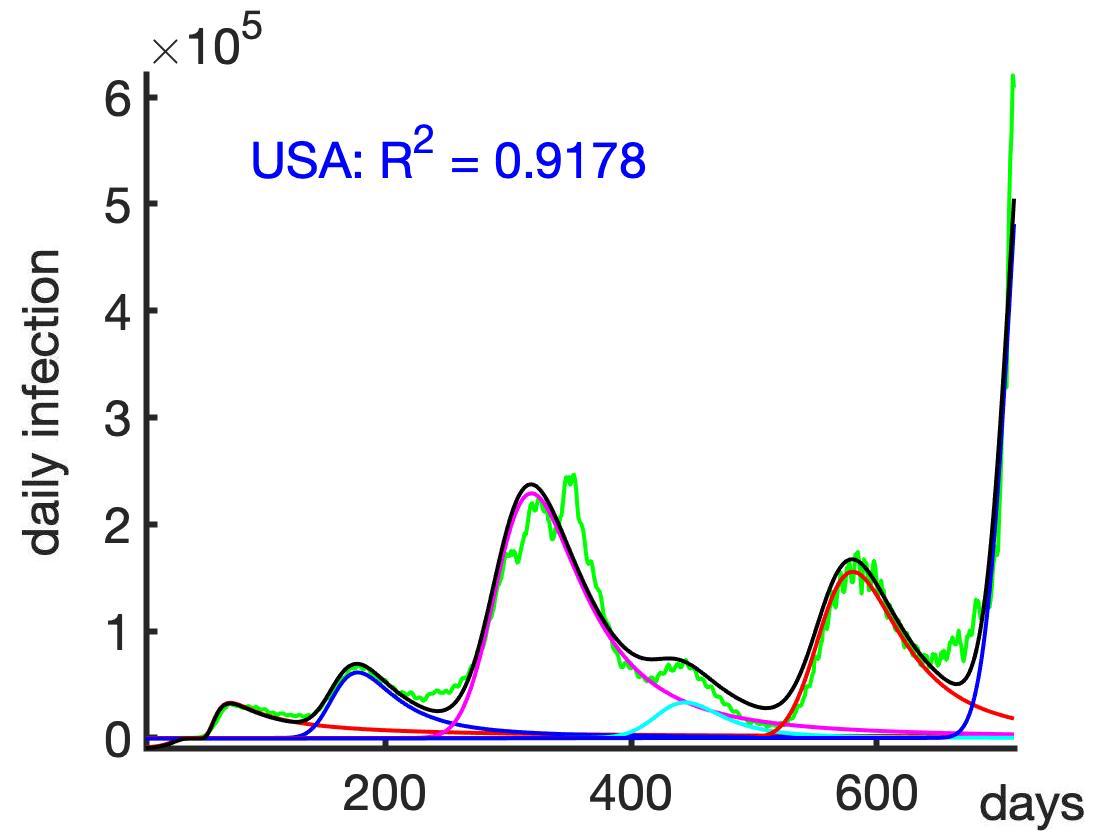}
\includegraphics[width=.47\textwidth]{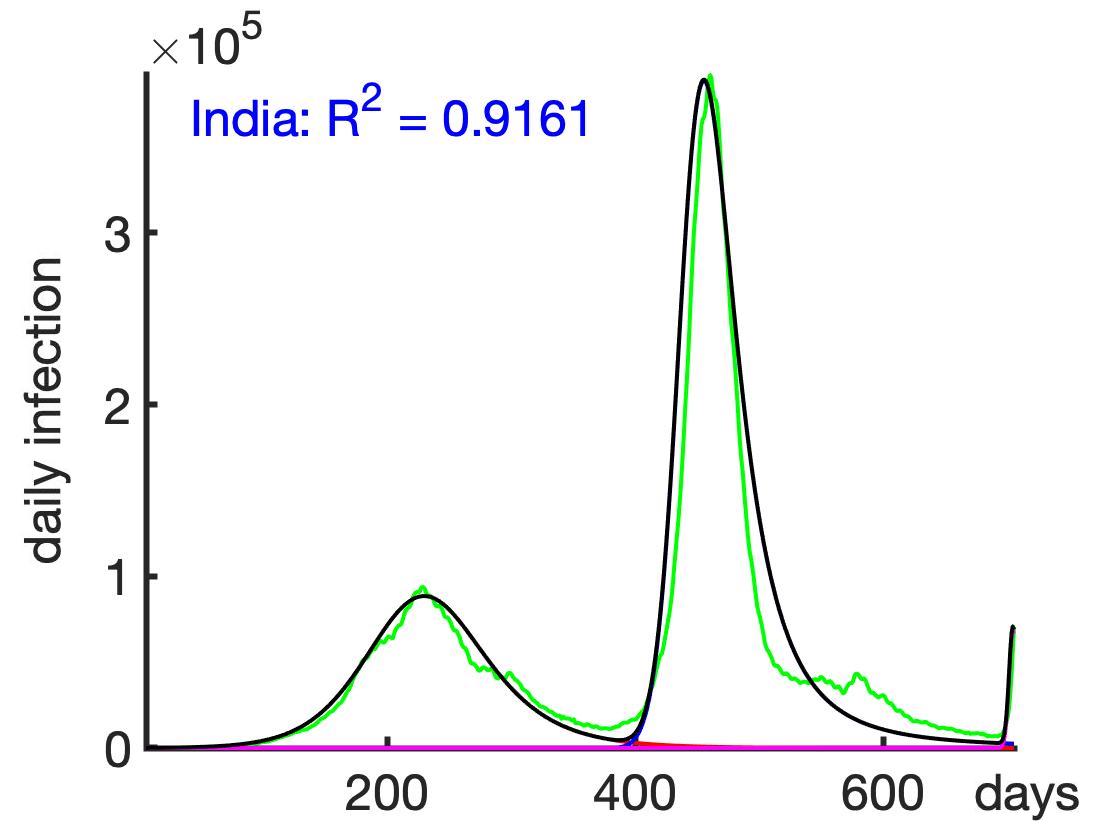}
\\
\includegraphics[width=.47\textwidth]{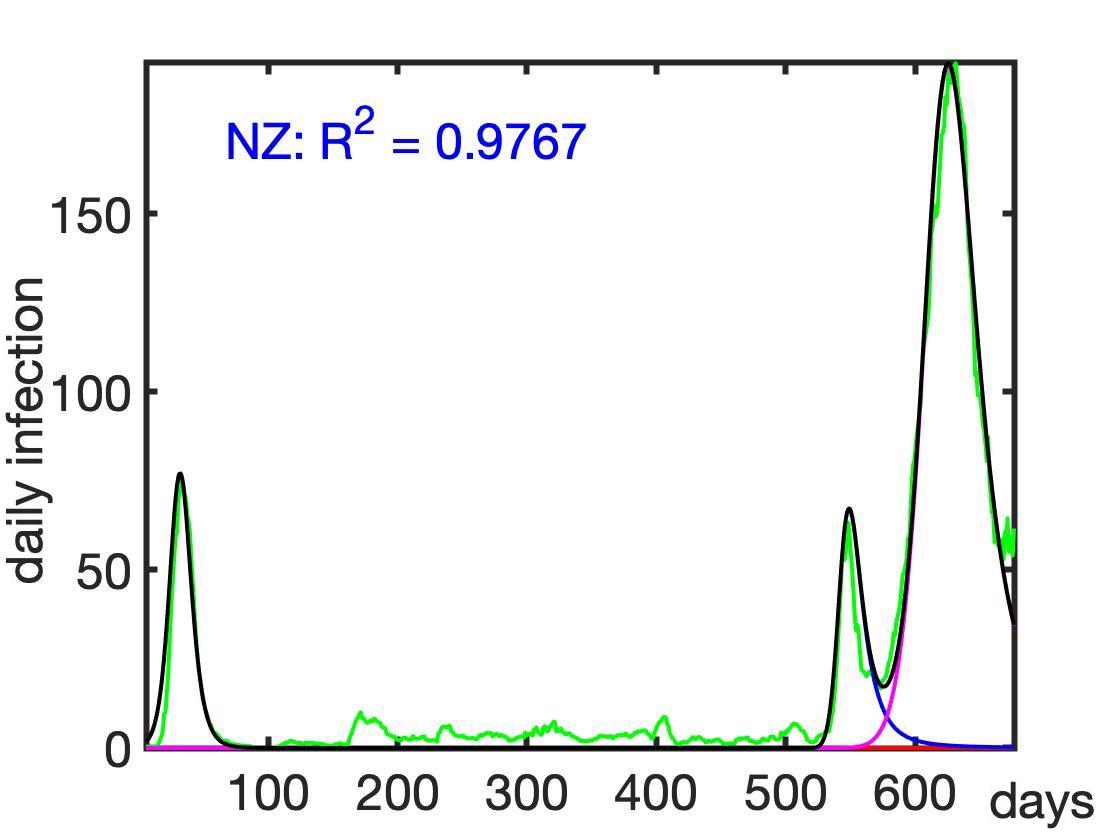}
\includegraphics[width=.47\textwidth]{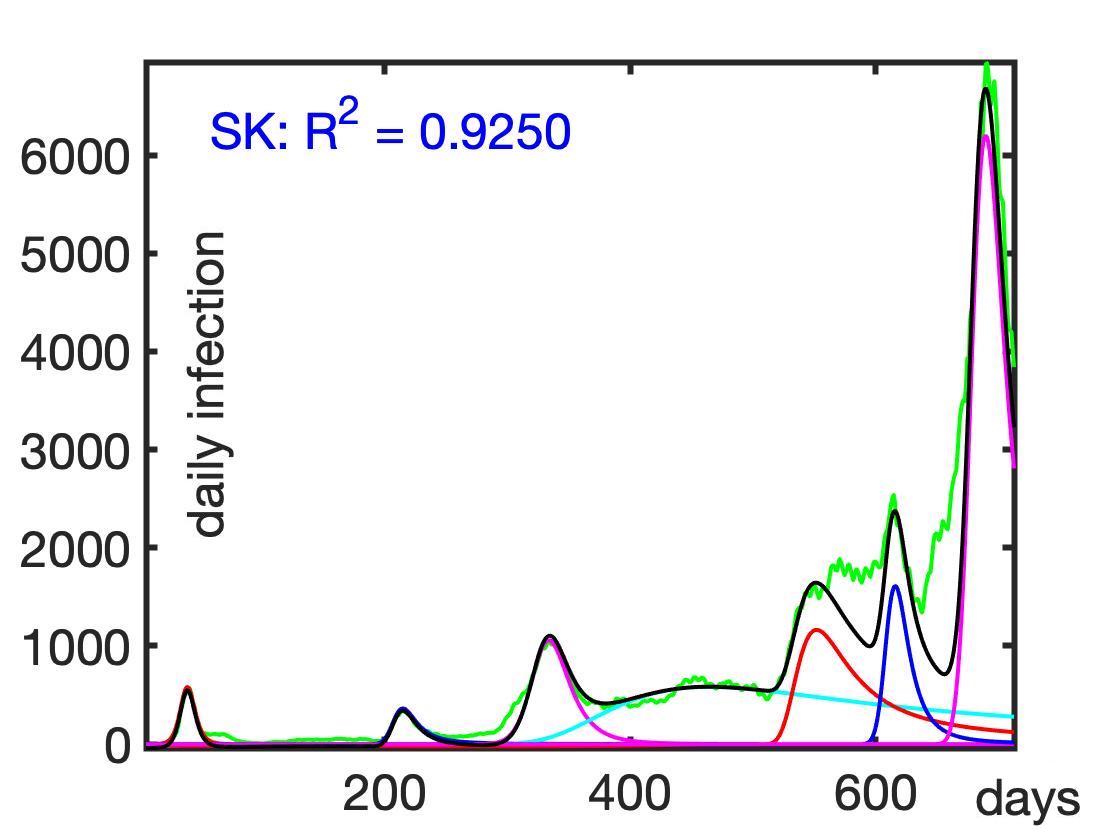}
\caption{{\it{Additive decomposition into multi-peaked data \cite{owid} using Eq.~\ref{eqadd}. Note that the $R^2$ values are still dominated by the largest peaks, but it is easier to fit valleys and extract those peaks with modest, potentially realistic, values for the mitigation parameter $\beta$. While our fits are not optimized for best-fitting algorithms, an ML/AI approach with the underlying fitting equations would generate more accurate estimates for these parameters. }} }
\label{figadd}
\end{figure*}
One way to address the valley effect is to assume that the recovered population goes back to being susceptible, giving us in effect, an SIS model and creating a robust residue of susceptibles for further infection, restoring the possibility of a moderate, physically meaningful $\beta_n$. 
\begin{eqnarray}
\dfrac{dS}{dt} &=& -ASI + BI\nonumber\\
\dfrac{dI}{dt} &=& ASI - BI
\label{eqsis}
\end{eqnarray}
assuming an instant reintroduction of a recovered population back into susceptible (we can also build delays as incubation periods/temporary immunity post infection). 
The solution to $s = S/N_p$ is straightforward. Once again, we differentiate the first equation with time, and substitute expressions from $f=I/N_p$ from the first equation and replace the derivative $df/dt = -ds/dt$ to get a differential equation involving $s$ alone. We then replace $d^2s/dt^2 = pdp/ds$, where $p=ds/dt$,  solve for $p$ using an integrating factor, and then solve the first order differential equation involving $p = ds/dt$, each time keeping track of initial conditions $p_0 = (\beta - s_0)f_0$, with $f_0 = 1-s_0$. The result is
\begin{equation}
    s(t) = \Biggl[\dfrac{f_0\beta +  (1-f_0-\beta)e^{-\displaystyle (1-\beta)t/\tau_0}}{f_0+ (1-f_0-\beta)e^{-\displaystyle (1-\beta)t/\tau_0}}\Biggr]
\end{equation}
where at $t=0$, $s = s_0 = 1 - f_0$. It is plain to see that for $\beta < 1$, $s$ approaches $s^* = \beta$ at long times, while for $\beta > 1$, $s$ approaches $1$, qualitatively consistent with the results of the single peak $SIR$ model in Fig.~\ref{f1}. Once again, we can interpret this evolution as an overdamped particle in a potential, except in this case (compare with Eq.~\ref{eqmotion})
\begin{equation}
    U_{SIS}(s) = \displaystyle 
    \dfrac{s^2(1-2s/3)}{2} + \beta\Biggl[\dfrac{s(s-2)}{2}\Biggr]
\end{equation}
with $AN_p = 1/\tau_0$, $\beta = B/AN_p$. As before, we are ignoring an overall constant vertical shift in $U(s)$ related to $s_0$ that has no bearings on the dynamics and amounts to choosing the (arbitrary) ground of the potential. 

There are many variants of the SIR model such as the SISV model \cite{sisv}, where a part of the susceptible population gets vaccinated, while a fraction of the vaccinated go back to being susceptible. Or the SIRS model, where the infected population gets split between a recovered population and a newly susceptible population.
There are other acronyms such as SIVR (SIR + virus variant), SIQR (SIR + Quarantine) \cite{SIAQN17}, SIAR (SIR + symptomatic vs asymptomatic), \cite{ASGS20}, SIR-S (SIR + stratification), SIXR (SIR + vaccination), P2SIR (SIR + travel) models \cite{connolly},\cite{connolly2}. They can also have added features such as deaths, maternally derived immunity, exposure period, etc. The result in the previous paragraph implies that in many cases we can identify a suitable hybrid between SIR and SIS models, which have different asymptotic behaviors $s^*_{SIR} \sim \beta^{2.5}$ and $s^*_{SIS} \sim \beta$. In fact, we can invoke a fraction that is re-inserted from the infected population back into susceptible (the rest to recovery), to fit an experimentally measured $s^*$ vs $\beta = 1/R_0$ in a controlled experimental environment.

Within the SIR model itself, one can avoid the issue with poor valley fitting by going back to an additive decomposition of the form
\begin{equation}
    \dfrac{df}{dt} = \sum_{i=1}^N \displaystyle\frac{N_{0i}e^{\displaystyle (t-t_{\beta_i})/\tau_{\beta_i}}}{\left(1+f_ie^{\displaystyle (t-t_{\beta_i})/\tau_{\beta_i}}\right)^2}, ~~~\tau_{\beta_i} = \tau_{0i} + \alpha_i(t-t_{\beta_i})
    \label{eqadd}
\end{equation}
Fig.~\ref{figadd} shows the impact of an additive fitted equation (Eq.~\ref{eqadd}) on the infection rate. We use $f_i = N_{0i}^{-1}$. The rest of the parameters are tabulated in Table 2. The calculated $R^2$ values are higher as shown in the figure, suggesting that we may get a better fit with an additive model. Further work will need to be done to connect these parameters with the mitigation parameters in the multiplicative model.
\begin{table}
\begin{math}
\begin{array}{ccccc}
\hline
& USA & India & New Zealand & S Korea\\
\hline
\tau_{01} & 1 & 22.8 & 4.8 & 3.3\\
\alpha_1 & 0.13 & 0.02 & 0.02 & 0.02\\
t_{b1} & 53 & 13 & 13 & 23\\
N_{01} & 800 & 3100 & 40 & 92\\
\tau_{02} & 3.5 & 4.3 & 3.3 & 2.6\\
\alpha_2 & 0.08 & 0.06 & 0.08 & 0.09\\
t_{b2} & 124 & 390 & 534 & 198\\
N_{02} & 1405 & 3600 & 35 & 78\\
\tau_{03} & 5 & 1 & 8 & 5.7\\
\alpha_3 & 0.07 & 0.055 & 0.055 & 0.05\\
t_{b3} & 234 & 695 & 580 & 295\\
N_{03} & 3205 & 650 & 90 & 180\\
\tau_{04} & 6.3 & N/A & N/A & 10.7\\
\alpha_4 & 0.06 & N/A & N/A & 0.13\\
t_{b4} & 370 & N/A & N/A & 320\\
N_{04} & 1200 & N/A & N/A & 280\\
\tau_{05} & 5 & N/A & N/A & 3\\
\alpha_5 & 0.07 & N/A & N/A & 0.12\\
t_{b5} & 500 & N/A & N/A & 520\\
N_{05} & 2600 & N/A & N/A & 185\\
\tau_{06} & 4 & N/A & N/A & 2.4\\
\alpha_6 & 0.07 & N/A & N/A & 0.08\\
t_{b6} & 660 & N/A & N/A & 597\\
N_{06} & 5600 & N/A & N/A & 160\\
\tau_{07} & N/A & N/A & N/A & 2.9\\
\alpha_7 & N/A & N/A & N/A & 0.08\\
t_{b7} & N/A & N/A & N/A & 660\\
N_{07} & N/A & N/A & N/A & 365\\
\hline
\end{array}
\end{math} 
\caption{Fitted data with an additive model (Eq.~\ref{eqadd}). The resulting plots fit the valleys better (Fig.~\ref{figadd}) with an $R^2$ number typically higher than the multiplicative version (Fig.~\ref{fig2}.)}
\label{tablet}
\end{table}

Note that standard device models for electron flow include a drift-diffusion component (drift is the sliding down the potential, and diffusion is an uncertainty related jitter that will be discussed shortly) and also a recombination-generation component. In this case, recombination would be a part of every SISV population that dies through natural causes and part of the infected population dying through infection related complications, while generation would be new births. Over the duration of a pandemic, spanning a few months, we can ignore birth and death rates and focus on a near constant population. 

We now discuss the sensitivity of the parameters, and overall dynamics of error propagation, which has implications both on long term predictability, and on effective strategies for frequency of data collection.

\section{Error Propagation}
While the above equations provide a simple fitting protocol for the spread of a pandemic, they do not carry any inherent predictive value as the relevant parameters are retrofitted. Predicting the parameters requires extensive data and insights into the underlying dynamics (e.g. linear vs nonlinear equations, time-dependence of parameters), typically both. To carry this forward we will need to generalize the SIR model to a spatio-temporal gradient diffusion equation, which is beyond the scope of this paper. 
While detailed epidemiological models can relate SIR parameters such as $\tau_0$, $\alpha$, $\beta_i$ etc to constants such as the reproduction number $R_0$ based on contact-tracing and cumulative incidence data, it is worth dwelling on the challenges of reliable prediction based on these numbers and equations alone. 

We identify three sources of error in our fitting protocols - (a) {\it{reporting error}} $\sigma_0$ which has to do with initial uncertainty in data collection (known unknowns), (b) {\it{parametric uncertainty}} $D$ which has to do with oversimplification in our evolution equations in the face of more complex and unpredictable microscopic and macroscopic interactions (unknown unknowns - governments enact lockdowns, a breakthrough happens in vaccine technology), as well as uncertainty in the parameters that evolve (known unknowns - e.g. virus mutates, people congregate at popular venues such as festivals), and (c) {\it{measurement error}} $(\epsilon, \delta)$ arising inherently from the finite sized and noisy nature of the data itself. Of these three, the first two belong to a common category ($D$ grows the initial uncertainty $\sigma_0$ linearly at first, later slowing it down to a sublinear function of time). 

\subsection{Reporting Error $\sigma_0$ and parametric uncertainty $D$}

\begin{figure}[h!]
\includegraphics[width=0.45\columnwidth]{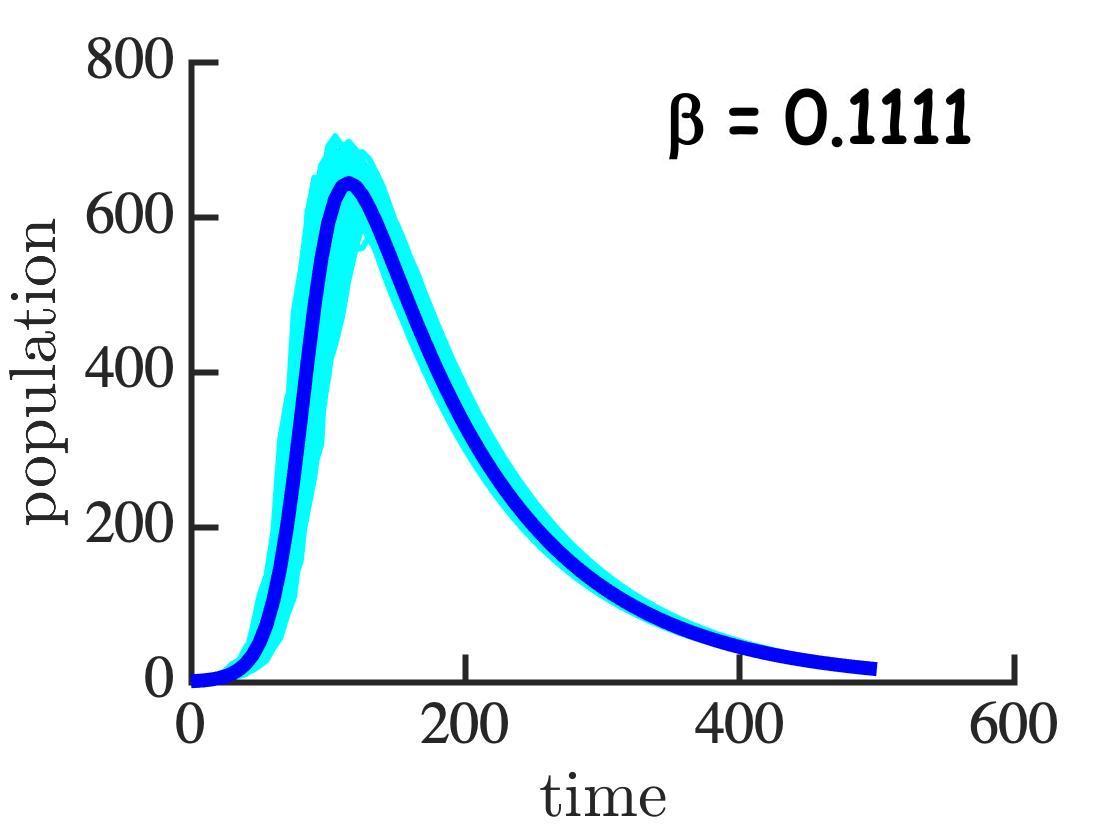}
\includegraphics[width=0.45\columnwidth]{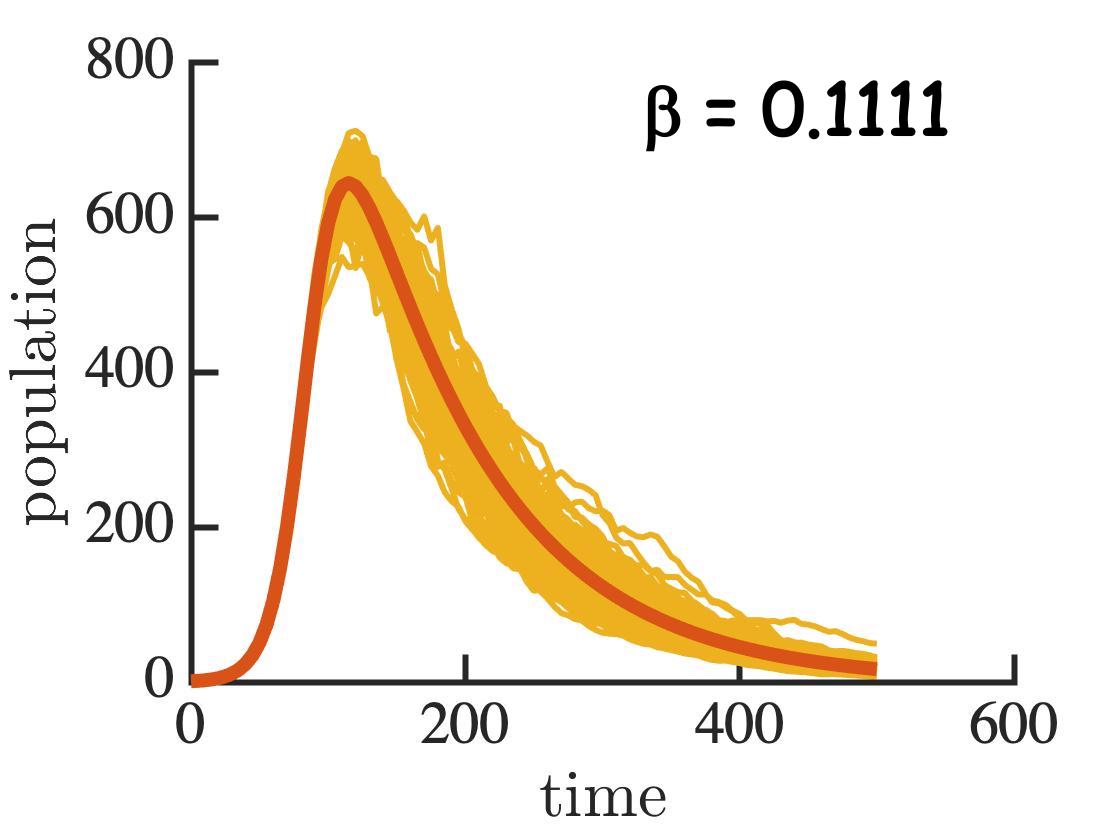}
\includegraphics[width=0.45\columnwidth]{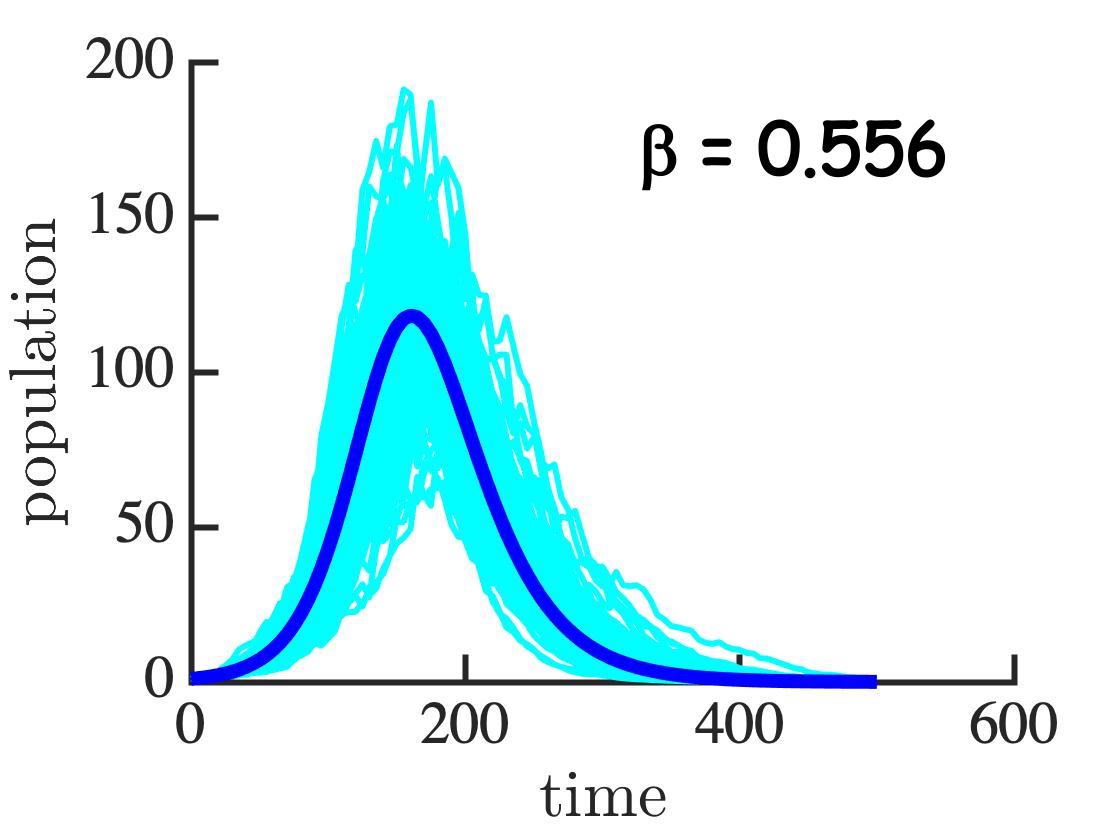}
\includegraphics[width=0.45\columnwidth]{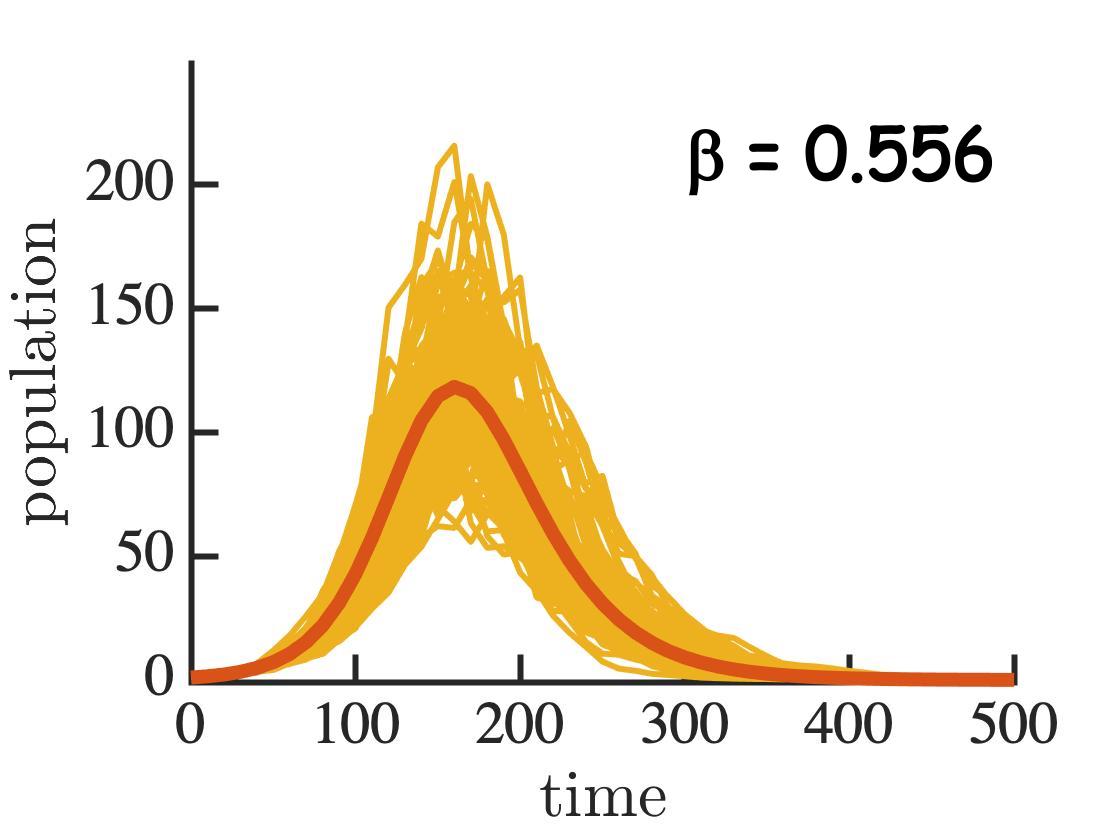}
\caption{{\it{(Top Row) {\it{Parametric Uncertainty.}} Results for a mean $\beta = 0.1111$ with a superposed random white noise on $A$ (left, 100 runs) and $B$ (right, 100 runs) with a noise of amplitude 25 population units. Here we assume $N_p = 1000$ and $\tau_0 = 11$ days. As expected, uncertainties in $A$ dominate the error before the peak and continue, while uncertainties in $B$ kick off after the peak.
(Bottom Row) Repeat for $\beta = 0.556$. For larger $\beta$ the noise is also larger, as is the uniformity of the uncertainty pre and post-peak, and the overall symmetry of the curves. }}}
\label{f9}
\end{figure}
Let us start by discussing how to add a random noise in the evolution of the probability distribution function (PDF). In presence of additive white noise $I_0$ the overdamped Newtonian evolution equation becomes the celebrated L\'angevin equation
\begin{equation}
    \displaystyle \frac{df}{dt^\prime} =   -\dfrac{1}{\tau_0}\frac{\partial U(f)}{\partial f} + I_0(t)
    \label{eqdiff1}
\end{equation}
with stretch $\alpha$ subsumed in $t^\prime$, where the noise has the following average moments
\begin{eqnarray}
    \langle I_0(t)\rangle &=& 0\nonumber\\ \langle I_0(t)I_0(t^\prime)\rangle &=& D\delta(t-t^\prime)
    \label{eqdiff2} 
\end{eqnarray}
where the diffusion constant $D$ is usually proportional to the mobility of the particle (velocity over force, set by the potential gradient and damping) and temperature which controls repulsive particle-particle interaction. 
The equation is generally solved using stochastic Monte Carlo techniques, where we use a random number generator to repeatedly construct $I_0$ values sampled from a given probability distribution. We then solve $f_0(t) = f(t,I_0)$ for each given $I_0$ (e.g. Fig. \ref{f9}) and extract a histogram. Indeed, this is one of the most popular ways of solving the pandemic equation, i.e., tossing coins to generate random values of $I_0$ from a given distribution and then numerically solving for $f$ - very often, this is done using a discretized (algebraic) deconstruction of the ODE onto a large grid of population elements and then the fraction $f$ is numerically extracted. 
It is however convenient, invoking the law of large numbers and ultimately the central limit theorem, to simplify the analysis (at least for intuitive reasoning) to directly estimate the PDF using the Fokker-Planck equation that comes from the L\'angevin equation. 
\\
Since $I_0$ is extracted from a probability distribution $\Pi(I_0)$, typically Gaussian white noise with variance $D$, the probability $P(f,t)$ for the output $f$ can be written as
\begin{equation}
    P(f,t) = \int dI_0\Pi(I_0)\delta(f-f_0(t))
\end{equation}
We can then calculate $dP/dt$, using the property of a Gaussian $I_0\Pi(I_0) = -D \partial \Pi/\partial I_0$ and the underlying Markov approximation, to derive the corresponding Fokker-Planck equation
\begin{eqnarray}
    \displaystyle   \frac{\partial  P}{\partial t} &=& -\frac{\partial J}{\partial f} \nonumber\\
    J &=& \displaystyle \dfrac{1}{\tau_0}P\frac{\partial U}{\partial f}
+ \frac{\partial(DP)}{\partial f}
\end{eqnarray}
where the first term on the right of the probability current density $J$ shows the deterministic drift of the PDF towards the local minima of $U$, while the second term shows the stochastic diffusion that tries to spread out and homogenize $P$ across the set of available $f$ values. 

For the steady-state solution ($t \rightarrow \infty, \partial P/\partial t = 0$), the value of $J$ is independent of $f$ (Kirchhoff's Law) and set by boundary conditions for a constant $D$. This solution is the Boltzmann equation of the form
\begin{equation}
    P_0(f) = P(f,t=\infty) = P_0e^{\displaystyle - U(f)/\tau_0D}
\end{equation}
and the initial distribution will show a combination of drift (sliding downhill and narrowing) and diffusion (spreading symmetrically and broadening) to transition over time until it maximizes at steady state to the value where $U$ is the lowest. 

The transient behavior of the Fokker-Planck equation is not easy to solve analytically, but we can account for its dominant components. For instance, if  we start with a Gaussian initial distribution $P(f,0) = \displaystyle Ce^{ \displaystyle -(f-f_0)^2/2\sigma_0^2}$, then solving the FPE in Fourier domain, we can show that over time it will tend to spread as
\begin{equation}
    P(f,t) = Ce^{\displaystyle -(f-f_0 - v_0t)^2/2(\sigma_0^2 + 2Dt)}
\end{equation}
where $v_0(f) = -(1/\tau_0)dU/df$ (this only works if we assume the potential varies slowly over the width of the PDF so that the linear expansion of $U$ around the peak of the PDF suffices). The distribution would 
move us back to $f=0$ if $B > A$ or to $f = 1$ if $B < A$ (Fig.~\ref{f3}). The constant $C$ must integrate to the total population, so that $C = N_p/\sqrt{2\pi \sigma_0^2}$. We can see in this solution both the drift component $v_0$ and the diffusion component $D$ playing their respective roles. 

This equation shows us two sources of error inherent in the system - the first is the initial {\it{reporting error}} $\sigma_0$ which originates at the outset, such as through faulty data gathering, reporting, testing inaccuracies etc. The second is the {\it{parametric uncertainty}}, characterized by $D$, where over time there is added uncertainty due to the very nature of pandemic spreading and our convoluted response to it. We can thereafter calculate the evolving mean 
\begin{equation}
    \langle f(t)\rangle = \int fP(f)df
\end{equation}
which should track the peak at $f_0 + v_0t$, but with a growing standard deviation $\sigma_0^2 + 2Dt$ that will make prediction harder beyond $t_{max} \approx (\epsilon^2-\sigma_0^2)/2D$, where $\epsilon$ is the maximum acceptable error in $f$. 

Let us now connect this uncertainty with the pandemic equation. The easiest way is to introduce a Gaussian white noise in the parameters, $A, B \rightarrow A, B +\lambda_{A,B} \eta(t)$, , where $\eta(t)$ is the normal probability distribution to mimick the noise ($\langle \eta(t)\eta(t^\prime)\rangle = \delta(t-t^\prime)$, meaning $\eta$ and $\lambda$ both have units of days$^{-1/2}$). For the SIR potential (Eq.~\ref{eqmotion}) this gives an added
stochastic force and a corresponding diffusion constant by mapping Eq.~\ref{eqmotion} to Eqs.~\ref{eqdiff1},~\ref{eqdiff2}.
\begin{equation}
D_A(s) = \dfrac{[\lambda_AN_p s(1-s)]^2}{2}, ~~D_B(s) = \dfrac{[\lambda_BN_p s\ln{s}]^2}{2}
\end{equation}
As expected, the uncertainties at the two fixed points $s= 0, 1$ are zero, seen also in Fig.~\ref{f9}.

Note that we assumed a Gaussian noise for simplicity, but that distribution has infinite support ($f$ values are unrestricted), while we need to operate within the range $(0,1)$ for $f$. For a tight standard deviation, and $\langle f\rangle $ lying between $\sim (\sigma,1-\sigma)$, with $\sigma$ being the standard deviation, we will for the most part see physically meaningful $f$ values, but on occasion we will see unphysical $f$s that venture out of this limit. We can either choose to eliminate those $f$ values, average over them, or switch to a uniform distribution over the range $(0,1)$, for which a physically intuitive Fokker-Planck equation, however, can be challenging to derive. 

\begin{figure}[h!]
    \includegraphics[width=0.45\columnwidth]{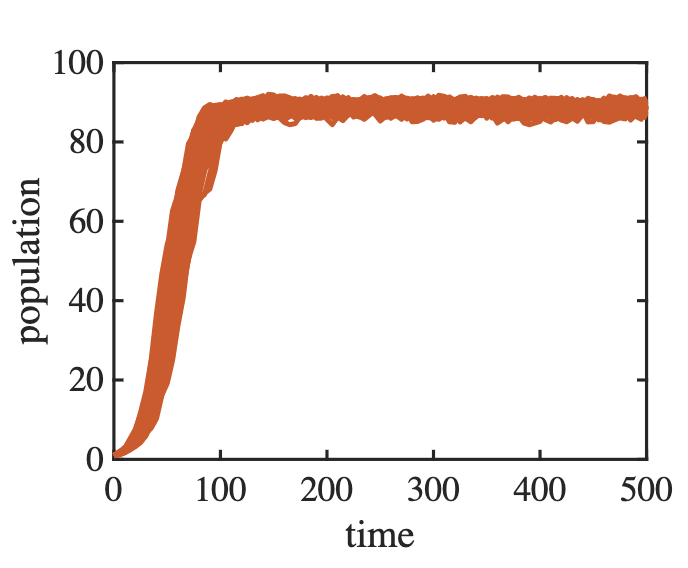}
    \includegraphics[width=0.5\columnwidth]{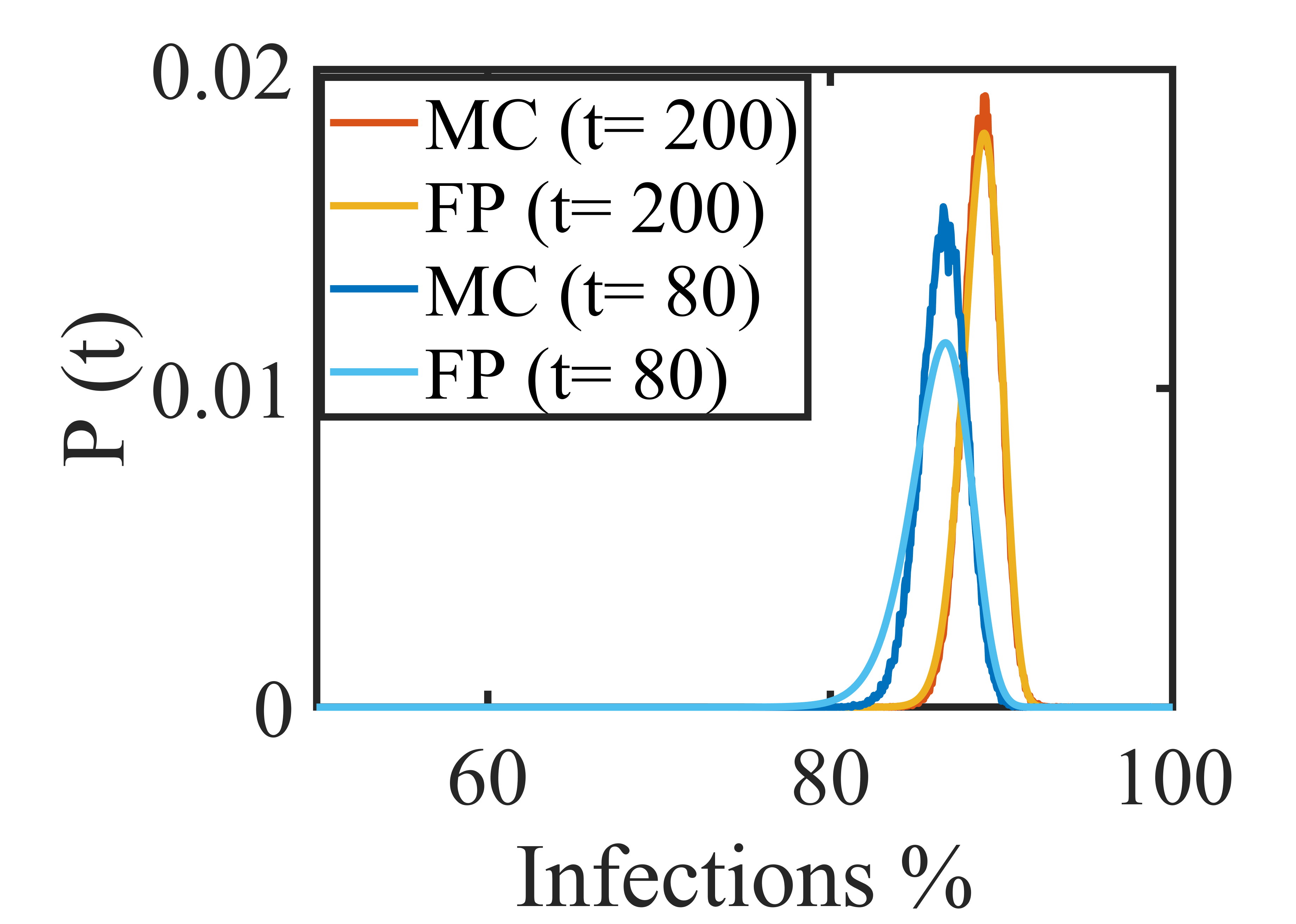}
    \caption{{\it{Left: I(t) is plotted against time. Which with out the recovery term it reaches an equilibrium at around 80. Due to the randomness in A, the final value has a range of error. Right: The Fokker-Planck solution compared with Monte Carlo (MC) simulation. Np is taken here to be 100. }}}
    \label{1DFP}
\end{figure}

In Fig.~\ref{1DFP}, we apply the Fokker-Planck (FP) equation to Eq.~\ref{eqsis} with noise in A. We take $A = A + \lambda\eta(t)$, where $\eta(t)$ is the normal probability distribution to mimick the noise. The Fokker Planck solution is compared to the Monte Carlo results with a random distribution for $A$. As can be seen from the figure, while diffusion tends to broaden the PDF, because the diffusion constant itself is $s$ dependent, there is a tightening of the distribution around the equilibrium value $s^*$, near which the diffusion constant closes to  $[\lambda N_p s^*(1-s^*)]^2/2$. This means that there is initially a growth in uncertainty but beyond peak infection that reduces as we reach the fixed point. 

As an illustration, suppose we have $\beta = 0.556$, and $\tau_0 = 10$ days the rise time for an infection. We begin with an initial reporting uncertainty $\sigma_0 = 0.1$ (i.e., $10\%$). We also assume a parameter uncertainty in $1/\tau_0 = AN_p$ equal to a fraction of $\Delta = 0.1$ ($10\%$ again). We can map this uncertainty with the standard deviation, meaning $\lambda^2 = \Delta/\tau_0$. 
The steady state $s^* = 0.27$, so the long time diffusion constant $D = [\lambda s^*(1-s^*)]^2/2 = 0.0002$/day. This means for the initial uncertainty to balloon up to say $\epsilon = 0.2$ (20 $\%$) will take $t_{max} \approx (\epsilon^2-\sigma_0^2)\tau_0/\Delta[s^*(1-s^*)]^2 \approx 75$ days (this analysis is admittedly over-simplified because we start from $s_0 \approx 1$ where the diffusion constant is also low, and the infection time $\tau$ has a time-dependent stretch that this back-of-the-envelope treatment ignores. However, we have outlined above the tools to calculate $s(t)$, $\tau(t)$ and do a more rigorous projection, should the need arise. Our estimated $D$ puts a lower bound on the data validity period, since $s(1-s)$ becomes maximum when $s$ reaches $0.5$ and $D \approx 0.0003$/day). 

For a multi-peaked solution, we can go a few steps further to estimate the time after which the Brownian particle can jump over the barriers of height $\Delta U$ in the $U(s)$ landscape, following an Arrhenius law $t_{jump}^{-1} = \nu e^{-\Delta U/\tau_0 D}$, where the attempt frequency $\nu$ is set by the dynamics in the valleys. Such a jump could transfer the configuration coordinates $s, f$ between metastable states (local minima) until subsequent noisy events can rescue them. We leave such analyses for future publications.

\subsection{Measurement uncertainty $\delta$}
\begin{figure}[ht!]
\includegraphics[width=0.475\columnwidth]{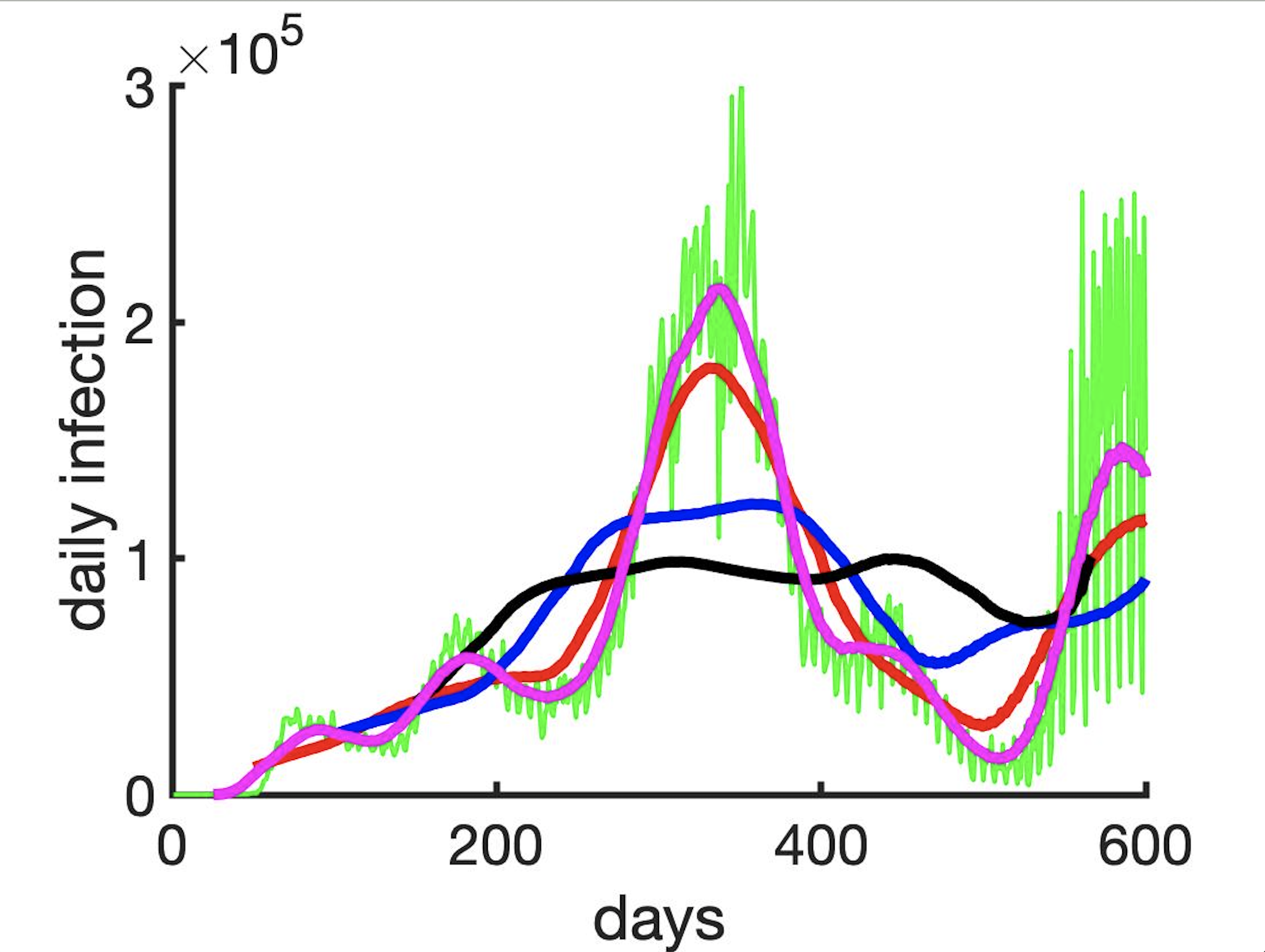}
\includegraphics[width=0.475\columnwidth]{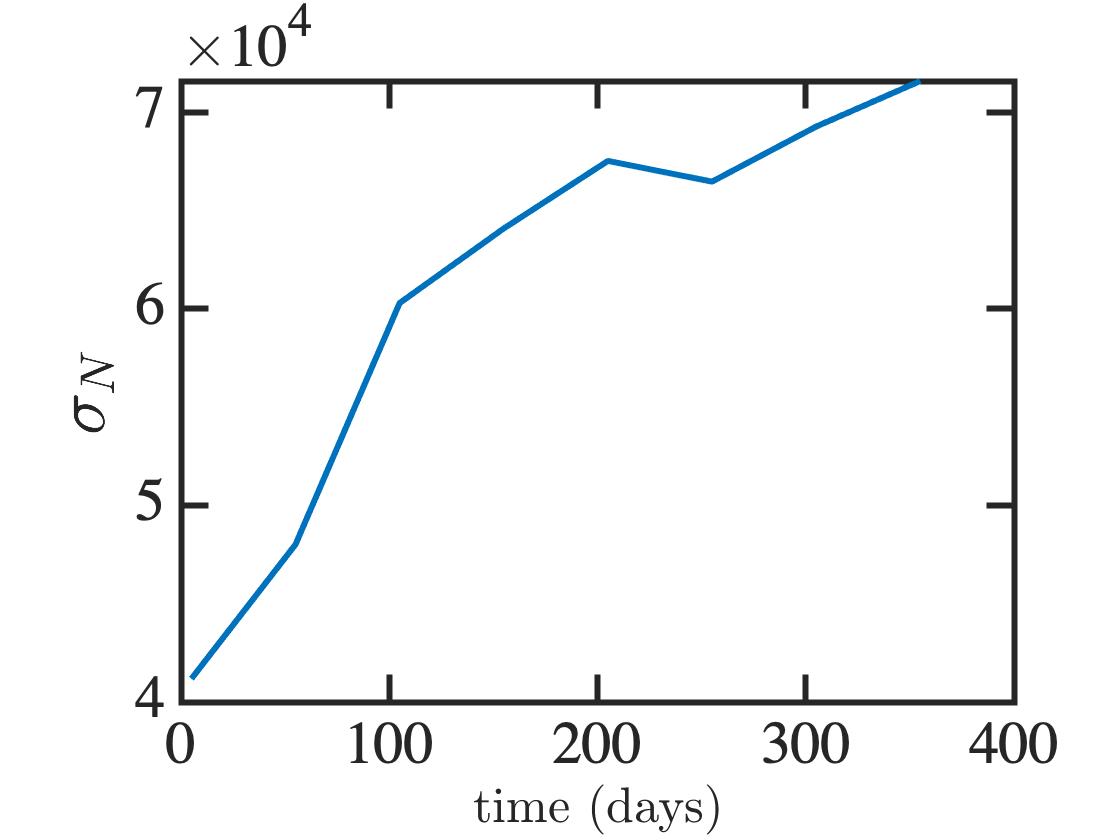}
\caption{{\it{(Left figure) Green: Raw daily data, magenta is $N_s$=50 ($N_s$ is the number of days over which a running average was done), red is $N_s$ = 100, blue is $N_s$ = 200 and black is $N_s$ = 300. (Right figure) Uncertainty grows like a Gaussian function with increasing sample size $N_s$ that averages out salient features.}}}
\label{f10}
\end{figure}

It is also worth emphasizing that there is an error with fitting the solutions to the Fokker-Planck equation to stochastic data over a finite dataset of sample size $N_s$. Based on Gaussian statistics, we can estimate that for an error margin $\epsilon$ (ie, accuracy probability $1-\epsilon$) the acceptable margin of error $(-\delta,\delta)$ around the running average for a finitely sampled set of size $N_s$ is given by  \cite{manohar}
\begin{equation}
\delta = 2\exp{[-2N_s\epsilon^2]}
\end{equation}
For a 5 day data period of averaging $N_s = 5$, we can say with 95$\%$ confidence that the data swing $\delta = 1.95$, meaning there is almost a $200 \%$ potential swing in smoothened data extracted, due to finite sample size errors. 
On the other hand, making the sampling $N_s$ too large has its problems, as that tends to average over and wash out the significant events. Fig.~\ref{f10} shows the error that builds up if the sampling time runs into 100s of days. Naturally, we expect an optimal sampling rate between these two limits.

\section{Conclusions}
Predicting pandemics is highly involved, as our knowledge of the underlying causes is often evolving in real time. Simple models provide broad insights, especially if we can find a way to visualize the evolution, develop phenomenological models with epidemiologically meaningful parameters, and have an accompanying error estimate. We have shown how we can visualize the pandemic response as the drift of an overdamped classical Brownian particle in a potential towards their local minima, along with uncertainty related diffusion away from those minima (strong enough uncertainty can diffuse the particle over a barrier into a  metastable state). The shape of the potential is controlled partly by intrinsic epidemiology, and partly by the various mitigation strategies and sociological constants at play. Simple multiplicative (Eq.~\ref{shurM}) and additive (Eq.~\ref{eqadd}) model fitting equations have been offered and their fits quantified across various countries. A part of the error arises from initial reporting uncertainty, which is amplified over time by parametric uncertainty except near fixed points where the parameters play minimal role on the dynamics. The quality of the data itself depends on the sampling time, where our ability to separate signal from noise (slower frequency evolution vs higher frequency random wiggles) poses a limit to the fitting equation and their overall predictability. 
With simple models and their uncertainties in place, we can focus on behavioral trends with respect to variation of these parameters such as cyclical vs abrupt quarantine measures. While these equations are the equivalent of a macrospin model in magnets (no spatial or geographical variation included), they need to eventually be extended to account for multi-patch solutions. Nonetheless, the quasi-analytical and easily visualizable cluster averages, and in particular error thresholds could be of potential use in predicting simple trends and evaluating policy decisions. 
\section{Acknowledgments}
We acknowledge initial discusions with Prof. Keith Williams (UVA, ECE) who suggested the use of Lotka-Volterra and SIR approaches, and later discusions with Prof. Anil Vullikanti (UVA, Biocomplexity institute). This project was funded by the UG internship program within the SRC-CRISP center, the NSF-REU supplement to the NSF-IUCRC center for Multifunctional Integrated Systems Technology (MIST), and the NSF grant CBET 1802588. 

\bibliography{apssamp}

\end{document}